\documentclass[nojss]{jss}

\usepackage{thumbpdf,lmodern}
\usepackage{bm}
\usepackage{amsthm,amsmath,amsfonts}
\usepackage{subcaption}

\author{Samuel W.K. Wong\\ University of Waterloo
   \And Shihao Yang\\ Georgia Institute of Technology 
   \And S.C. Kou \\ Harvard University}
\Plainauthor{Samuel W.K. Wong, Shihao Yang, S.C. Kou}

\title{\pkg{MAGI}: A Package for Inference of Dynamic Systems from Noisy and Sparse Data via Manifold-constrained Gaussian Processes}
\Plaintitle{MAGI: A Package for Inference of Dynamic Systems from Noisy and Sparse Data via Manifold-constrained Gaussian Processes}
\Shorttitle{\pkg{MAGI}: MAnifold-constrained Gaussian process Inference}

\Abstract{
This article presents the \pkg{MAGI} software package for the inference of dynamic systems. The focus of \pkg{MAGI} is on dynamics modeled by nonlinear ordinary differential equations with unknown parameters. While such models are widely used in science and engineering, the available experimental data for parameter estimation may be noisy and sparse. Furthermore, some system components may be entirely unobserved. \pkg{MAGI} solves this inference problem with the help of manifold-constrained Gaussian processes within a Bayesian statistical framework, whereas unobserved components have posed a significant challenge for existing software. We use several realistic examples to illustrate the functionality of \pkg{MAGI}. The user may choose to use the package in any of the \proglang{R}, \proglang{MATLAB}, and \proglang{Python} environments.}


\Address{
	Samuel W.K. Wong \\ 
	Department of Statistics and Actuarial Science \\
	University of Waterloo \\
	200 University Ave W \\
	Waterloo, ON N2L 3G1, Canada \\
	E-mail: \email{samuel.wong@uwaterloo.ca} \\

	Shihao Yang \\
	H. Milton Stewart School of Industrial and Systems Engineering \\
	Georgia Institute of Technology \\
	755 Ferst Drive NW \\
	Atlanta, GA 30332, USA \\
	E-mail: \email{shihao.yang@isye.gatech.edu} \\

	S.C. Kou \\
	Department of Statistics \\
	Harvard University \\
	1 Oxford St, 7th floor \\
	Cambridge, MA 02138, USA \\
	E-mail: \email{kou@stat.harvard.edu}
}

\begin{document}

\section{Introduction} \label{sec:intro}

Ordinary differential equations (ODEs) are widely used as models for dynamic systems in science and engineering, including gene regulation \citep{bolouri2008computational}, chemical reactions \citep{walas1991modeling,wong2023estimating}, epidemiology and ecology \citep{busenberg2012differential}, economics \citep{tu2012dynamical}, etc. We focus here on the case where the ODEs are nonlinear with unknown parameters governing their behavior. The problem of interest is to recover the unobserved system trajectories as well as to estimate the parameters from experimental or observational data, where the observations taken from the system may be subject to measurement noise and may only be available at a sparse number of time points.  Further, some components in the system may be entirely unobserved. This paper introduces the \pkg{MAGI} software package, named after its corresponding method \citep[MAnifold-constrained Gaussian process Inference;][]{yang2021inference} which provided fast and accurate inference for this statistical problem on a variety of examples, including the case when there are unobserved system components.

Specifically, \pkg{MAGI} handles parameter estimation for models where the system components are governed by a set of ODEs, which we denote by
\begin{equation}\label{eq:ode}
\dot{\bm{x}}(t) = \frac{d \bm{x}(t)}{dt} = \mathbf{f}(\bm{x}(t),\bm{\theta}, t), 
\end{equation}
where $\bm{x}(t)$ is the $D$-dimensional system trajectory over time $0 \le t \le T$ (i.e., $\bm{x} : [0,T] \to \mathbb{R}^D$), and $\dot{\bm{x}}(t)$ is shorthand for the vector of derivatives $d \bm{x}(t)/dt$, which are specified via the known function $\mathbf{f}$.  The vector $\bm{\theta}$ denotes the model parameters to be estimated, which govern the behavior of the system. We let $\bm{y}(\bm{\tau})$ denote the observed data, namely the noisy measurements taken from the system at observation time points $\bm{\tau}$. Throughout this article, we use $\bm{\tau} = (\bm\tau_1, \bm\tau_2, \dots, \bm\tau_D)$ to denote the collection of observation time points, where $\bm\tau_d$ is the vector of time points at which component $d$ is observed, $d=1,\ldots, D$. Each system component can have its own set of observation times $\bm\tau_d$, and some components may not be observed at all (for which $\bm \tau_d = \emptyset$). We assume that the noise is additive and Gaussian, i.e., $\bm{y}(\bm{\tau}) = \bm{x}(\bm{\tau}) + \bm{\epsilon}(\bm{\tau})$, where the error term $\bm{\epsilon}$ has noise level $\bm{\sigma}$ (which may be known or unknown). The key feature of \pkg{MAGI} is to infer $\bm{x}(t)$ and $\bm{\theta}$ from $\bm{y}(\bm{\tau})$ without the need for any numerical integration, even when there are unobserved system components.\footnote{In practice, \pkg{MAGI} infers $\bm{x}(t)$ for all $t \in \bm{I}$, where $\bm{I}$ is any finite set of discretization points in $[0,T]$ specified by the user, as subsequently demonstrated.} This is achieved by taking a Gaussian process (GP) as a prior for $\bm{x}(t)$ and constraining it to a manifold that satisfies the ODE system. Inference is then carried out within a principled Bayesian statistical framework, that is, we condition on all known information and quantities and apply Bayesian techniques to the resulting posterior distribution.

\pkg{MAGI} is available for \proglang{R}, \proglang{MATLAB}, and \proglang{Python} which enable practitioners to input and work with custom ODE systems in their preferred computing environment.  (The packages share a common \proglang{C++} codebase, which ensures a consistent method implementation across all three environments.)  The main text of this article will provide code examples in \proglang{R} \citep{R}. Equivalent code for the examples in \proglang{MATLAB} and \proglang{Python} are provided in the replication materials, along with usage instructions in the Appendices \ref{app:matlab} and \ref{app:python}. The replication materials can be downloaded from Github: \url{https://github.com/wongswk/magi/tree/master/JSS-replication}.

\subsection{Illustrative example: oscillation of Hes1 mRNA and protein levels}\label{sec:hes1intro}

To begin with a concrete example, consider the three-component dynamic system, which for short we write as $X = (P,M,H)$, introduced in \citet{hirata2002oscillatory}, governed by the ODEs
\begin{align}
\mathbf{f}(X, \bm{\theta}, t) = \begin{pmatrix}
-aPH + bM - cP \\
-dM + \frac{e}{1 + P^2} \\
-aPH + \frac{f}{1+ P^2} - gH
\end{pmatrix}, \label{eq:hes1nolog}
\end{align}
where $P$ and $M$ are the protein and messenger ribonucleic acid (mRNA) levels in cultured cells.  In experimental data, $P$ and $M$ levels exhibit oscillatory cycles approximately every 2 hours, and the $H$ component is a Hes1-interacting factor that helps regulate this oscillation via a negative feedback loop.  The parameters of this system are $\bm{\theta} = (a, b, c, d, e, f, g)$, where $a$ and $b$ can be interpreted as synthesis rates; $c,d$ and $g$ as decomposition rates; and $e$ and $f$ as inhibition rates.

The remainder of this subsection describes a realistic sample dataset that is simulated from this system. The key features of the dataset are: (i) $M$ and $P$ are measured at different sets of time points, (ii) the observations for $M$ and $P$ are noisy (i.e., have measurement error), (iii) $H$ is never observed.  In Section \ref{sec:basicusage}, we will then demonstrate how to use \pkg{MAGI} to recover the system trajectories and parameters from the dataset, without the use of any numerical solvers.

We first define a function that computes $\mathbf{f}$ for this ODE system. Its inputs are a vector of parameters $\bm{\theta}$, a matrix for $X$ (with columns corresponding to components), and a vector of time points \code{tvec}. The function then returns a matrix of values of $\mathbf{f}$, with rows corresponding to the time points in \code{tvec} and columns corresponding to the components of $X$: 
\begin{Sinput}
R> hes1modelODE <- function(theta, x, tvec) {
+   P = x[, 1]
+   M = x[, 2]
+   H = x[, 3] 
+   
+   PMHdt = array(0, c(nrow(x), ncol(x)))
+   PMHdt[, 1] = -theta[1] * P * H + theta[2] * M - theta[3] * P
+   PMHdt[, 2] = -theta[4] * M + theta[5] / (1 + P^2)
+   PMHdt[, 3] = -theta[1] * P * H + theta[6] / (1 + P^2) - theta[7] * H
+   
+   PMHdt
+ }
\end{Sinput}

Following the real experimental setup in \citet{hirata2002oscillatory}, measurements of $P$ and $M$ are taken every 15 minutes over a four-hour period, but asynchronously:  $P$ is observed at $t = 0, 15, 30, \ldots, 240$ minutes, while $M$ is observed at $t = 7.5, 22.5, \ldots, 232.5$ minutes, and $H$ is never observed.

We shall simulate from this system using the parameter values studied theoretically in \citet{hirata2002oscillatory}: $a = 0.022$, $b = 0.3$, $c = 0.031$, $d = 0.028$, $e = 0.5$, $f = 20$,  $g = 0.3$; the initial conditions $P(0) = 1.439$, $M(0) = 2.037$, $H(0) = 17.904$ are taken to mimic the authors' setting, where the system is initialized at the point in the stable oscillation cycle where $P$ is at its minimum. The observation noise in the experiment is approximately 15\% of both the $P$ and $M$ levels, which we treat as multiplicative noise following a log-normal distribution with known standard deviation 0.15. For convenience, we setup a list containing these input values for the simulation:
\begin{Sinput}
R> param.true <- list(
+   theta = c(0.022, 0.3, 0.031, 0.028, 0.5, 20, 0.3),
+   x0 = c(1.439, 2.037, 17.904),
+   sigma = c(0.15, 0.15, NA))
\end{Sinput}
Since $H$ is never observed, measurement noise is not applicable to that component. Next, to simulate the data for our analysis, we use a numerical solver to construct the system trajectories implied by these values. In \proglang{R}, we can utilize the ODE solvers available in the  \pkg{deSolve} package \citep{deSolve_Rpack}, by defining a wrapper that satisfies the syntax of its \code{ode} function. We emphasize that the numerical ODE solver is only used here for generating the data; throughout the \pkg{MAGI} method for inferring the system trajectories and parameters, no numerical solver is ever needed.
\begin{Sinput}
R> modelODE <- function(tvec, state, parameters) {
+    list(as.vector(hes1modelODE(parameters, t(state), tvec)))
+  }
\end{Sinput}

We may now numerically solve the ODE trajectories \code{x} over the time period of interest, from $t=0$ to $4$ hours (specified as 240 minutes):
\begin{Sinput}
R> x <- deSolve::ode(y = param.true$x0, times = seq(0, 60 * 4, by = 0.01),
+                   func = modelODE, parms = param.true$theta)
\end{Sinput}

Next, we extract the true values of the trajectory at the time points according to the schedule of observations described, and simulate noisy measurements \code{y} for $P$ and $M$.  The seed \code{12321} is set for reproducibility. 
\begin{Sinput}
R> set.seed(12321)
R> y <- as.data.frame(x[ x[, "time"] %in% seq(0, 240, by = 7.5), ])
R> names(y) <- c("time", "P", "M", "H")
R> y$P <- y$P * exp(rnorm(nrow(y), sd = param.true$sigma[1]))
R> y$M <- y$M * exp(rnorm(nrow(y), sd = param.true$sigma[2]))
\end{Sinput}
For system components that are unobserved at a time point, we fill in the corresponding  values with \code{NaN}, recalling that $P$ and $M$ are observed asynchronously and $H$ is never observed:
\begin{Sinput}
R> y$H <- NaN
R> y$P[y$time %in% seq(7.5, 240, by = 15)] <- NaN
R> y$M[y$time %in% seq(0, 240, by = 15)] <- NaN
\end{Sinput}

\begin{figure}[ht]
	\centering
	\includegraphics[width=5.5in]{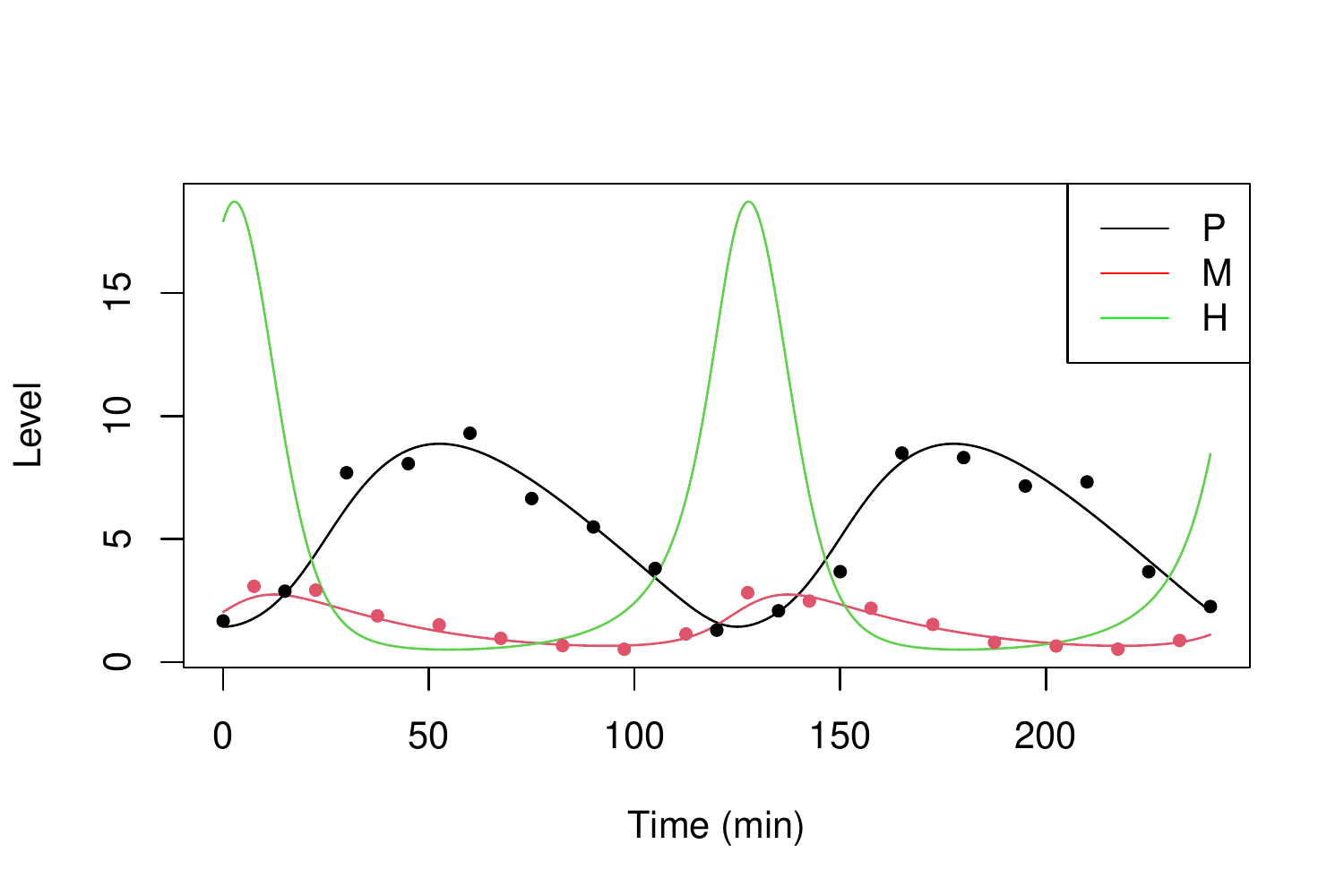}
	\caption{True system trajectories (solid curves) and sample noisy observations (points) from the Hes1 system.  The $H$ component is never observed. In Section \ref{sec:basicusage}, we demonstrate how \pkg{MAGI} recovers the system trajectories and parameters from this sample dataset of noisy observations.} \label{fig:hes1obs}
\end{figure}

Now the dataset \code{y} is prepared.  Based on this dataset, \pkg{MAGI} will infer the underlying trajectories $X$ and estimate the seven parameters in $\bm{\theta}$.   We plot the observed data in Figure \ref{fig:hes1obs}, with the points showing the noisy measurements available for $P$ and $M$ and the solid curves showing the true system trajectories of $X$.  The following commands create this plot:
\begin{Sinput}
R> compnames <- c("P", "M", "H")
R> matplot(x[, "time"], x[, -1], type = "l", lty = 1, 
+         xlab = "Time (min)", ylab = "Level")
R> matplot(y$time, y[,-1], type = "p", col = 1:(ncol(y)-1), pch = 20, add = TRUE)
R> legend("topright", compnames, lty = 1, col = c("black", "red", "green"))
\end{Sinput}

\subsection{Overview of related software}\label{sec:related}

\pkg{MAGI} handles the so-called \emph{inverse problem} for ODEs, namely to recover the system trajectories and parameters from a set of observational or experimental data. Existing software for this problem can be broadly categorized into methods that rely on using numerical solvers for ODEs and those that do not.  

In Equation~\ref{eq:ode}, the system trajectory $\bm{x}(t)$ may be determined for a given set of parameter values $\bm{\theta}$ and initial conditions $\bm{x}(0)$ by integration. For nonlinear functions $\mathbf{f}$, numerical integrators (e.g., Runge-Kutta) are often needed for solving the ODEs in this way. We may denote this numerical solution as $\hat{\bm{x}}(t; \bm{x}(0), \bm{\theta})$ to indicate its deterministic relationship with $\bm{\theta}$ and $\bm{x}(0)$. A simple approach can thus repeatedly solve for $\hat{\bm{x}}(t; \bm{x}(0), \bm{\theta})$ to optimize a likelihood or least-squares criterion for the data $\bm{y}(\bm{\tau})$, as a function of $\bm{\theta}$ (and also $\bm{x}(0)$ if the initial conditions are unknown).
A least-squares criterion to minimize would take the form $\left\Vert \bm{y}(\bm{\tau}) - \hat{\bm{x}}(\bm{\tau}; \bm{x}(0), \bm{\theta}) \right\Vert^2$ where $\left\Vert \cdot \right\Vert$ is the usual Euclidean norm, i.e., by evaluating $\hat{\bm{x}}$ (which is determined for all $t$) at the observation time points $\bm{\tau}$.
This can be viewed as a non-linear least squares (NLS) problem, which could be handled in \proglang{R} using \code{nls} from \pkg{stats} \citep{R} together with one of the numerical integrators in \pkg{deSolve}. In \proglang{MATLAB}, the \pkg{System Identification} toolbox \citep{ljung1995system} and the \pkg{Data2Dynamics} modeling environment \citep{raue2015data2dynamics} also provide functionality for inverse problems with the help of numerical solvers.  Bayesian approaches for parameter estimation using numerical solvers are also available, such as in the \pkg{deBInfer} package  \citep{boersch2017debinfer} in \proglang{R}. In this case, the numerical solution is used to construct a likelihood $p(\bm{y}(\bm{\tau}) | \hat{\bm{x}}(\bm \tau; \bm{\theta}, \bm{x}(0) ), \bm{\sigma})$ and priors are placed on all the unknown quantities among $\bm{\theta}$, $\bm{x}(0)$, and $\bm{\sigma}$.
While methods based on numerical solvers are generally applicable for ODE inverse problems (including when system components are unobserved), they may encounter significant computational bottlenecks: the numerical solver must be invoked repeatedly for values of $\bm{\theta}$ and $\bm{x}(0)$ used in the estimation procedure.

As a result, the second broad category consists of methods designed to estimate $\bm{\theta}$ without the need for numerical integration; \pkg{MAGI} belongs to this category. These methods use various techniques to curve-fit the observations while following the ODE system dynamics (e.g., by gradient matching). We overview some representative methods with software available in \proglang{R} in the following:
\begin{itemize}
\item A pioneering collocation approach proposed a B-spline basis to fit the system trajectories, where $\bm{x}(t)$ is represented by $\hat{\bm{x}}(t)= \bm{c}^\top \bm{\Phi}(t)$ for basis functions $\bm{\Phi}(t)$ and coefficients $\bm{c}$. Then, a penalized likelihood of the form $\left\Vert \bm{y}(\bm{\tau}) - \hat{\bm{x}}(\bm{\tau}) \right\Vert^2 + \lambda \int [\dot {\hat{\bm{x}}}(t) -  \mathbf{f}(\hat{\bm{x}}(t),\bm{\theta}, t)  ]^2 \,dt$ is optimized, where the first term measures fit to the data and the second term (with penalty parameter $\lambda$ and using B-spline derivatives for $\dot {\hat{\bm{x}}}(t)$) ensures fidelity to the ODEs \citep{ramsay2007parameter}.  This method is available in the packages \pkg{CollocInfer} \citep{hooker2016collocinfer} and \pkg{pCODE} \citep{pCODE_Rpack}.
\item Reproducing kernel Hilbert spaces (RKHS) have also been used for gradient matching \citep{niu2016fast}. The $d$-th component of $\bm{x}(t)$ is represented by $\hat{x}_d(t)= 
 \bm{b}_d^\top \bm{k}_d(t)$ with $\bm{k}_d(t) = [k(t, t_1), \ldots, k(t,t_n)]$, where $k(t, \cdot)$ is a kernel function from the Hilbert space, $t_1,\ldots, t_n$ are the observation times, and $\bm{b}_d$ are kernel coefficients. A criterion of the form $\left\Vert \bm{y}(\bm{\tau}) - \hat{\bm{x}}(\bm{\tau}) \right\Vert^2 + \lambda \left\Vert \dot {\hat{\bm{x}}}(\bm{\tau}) -  \mathbf{f}(\hat{\bm{x}}(\bm{\tau}),\bm{\theta}, \bm{\tau})  \right\Vert ^2$ is then minimized, where the two terms have similar interpretation as in \pkg{CollocInfer}, and the regularization parameter $\lambda$ may be obtained by cross-validation.
 Different variants of RKHS methods, including transformations on $t$ to better accommodate potential time inhomogeneity of the ODE solutions, are implemented in \pkg{KGode} \citep{niu2021r}.
\item Inference based on a separable integral-matching approach is implemented in \pkg{simode} \citep{simode_Rpack}. First, ${\bm{x}}(t)$ is approximated via fitting a spline representation $\hat{\bm{x}}(t)$ to the data, to bypass numerical integration of the ODEs. Noting that the true ODE solution may be expressed as $\bm{x}(t) = \bm{x}(0) + \int_0^t \mathbf{f}({\bm{x}}(s),\bm{\theta}, s) ds $, integral matching then seeks to minimize the criterion $ \int_0^T \left\Vert \hat{\bm{x}}(t) - \bm{x}(0) -  \int_0^t \mathbf{f}(\hat{\bm{x}}(s),\bm{\theta}, s) \, ds \right\Vert^2 \, dt$ as a function of $\bm{\theta}$ and $\bm{x}(0)$. In systems where $\mathbf{f}$ is linear or semi-linear in $\bm{\theta}$, the estimates can be obtained efficiently by taking advantage of the separable parameters in the optimization procedure (otherwise, non-linear optimization will be required).
\item A Bayesian approach using GPs for fitting the trajectories is available in the \pkg{deGradInfer} package \citep{deGrad_Rpack}. This implements the gradient matching method of \citet{dondelinger2013ode}, which places a GP prior on $\bm{x}(t)$ (with hyper-parameters $\bm{\phi}$) so that $\bm{y}, \bm{x}, \dot{\bm{x}}$ have a joint GP specification. The joint distribution over all the quantities, namely $p(\bm{y}, \bm{x}, \dot{\bm{x}},\bm{\theta},\bm \phi, \bm{\sigma})$, is then factorized as $p(\bm{y}, \bm{x}, \dot{\bm{x}},\bm{\theta},\bm \phi, \bm{\sigma}) = p( \bm{y} | \bm{x}, \bm \sigma) p(\dot{\bm{x}} | \bm{x}, \bm{\theta}, \bm \phi)  p(\bm{x} | \bm \phi)  p( \bm \phi ) p( \bm{\theta} )p( \bm{\sigma} )$, where $p( \bm{y} | \bm{x}, \bm \sigma)$ denotes the likelihood of the observations, $p(\bm{x} | \bm \phi)$ is the GP prior on $\bm{x}(t)$, and priors are assigned to $\bm\theta, \bm\phi, \bm\sigma$. To carry out inference in this method, the term $p(\dot{\bm{x}} | \bm{x}, \bm{\theta}, \phi)$ is expressed as a heuristic product that combines the contributions of the GP and ODEs, namely $p(\dot{\bm{x}} | \bm{x}, \bm{\theta}, \phi) \propto p(\dot{\bm{x}} | \bm{x}, \phi) p(\dot{\bm{x}} | \bm{x}, \bm{\theta})$, where $p(\dot{\bm{x}} | \bm{x}, \phi)$ comes from the GP and $p(\dot{\bm{x}} | \bm{x}, \bm{\theta})$ comes from the specification of $\mathbf{f}$ in the ODE (see Equation~\ref{eq:ode}).
\end{itemize}
While some of these methods can handle unobserved system components in theory, the available software implementations tend to lack this functionality in general. Only \pkg{CollocInfer} and \pkg{pCODE} can accommodate an unobserved component in the estimation procedure; however, substantial manual input is required to carry out the analysis. (This is subsequently demonstrated in Section \ref{sec:comparisons}, where we carry out an illustrative comparison between methods.) Thus, one distinct contribution of the \pkg{MAGI} package is that it provides a ready-made solution for systems with unobserved components, in addition to its principled inference framework that is rooted in Bayesian statistics.

\section{Manifold-constrained Gaussian process inference}

This section explains the key points of the \pkg{MAGI} method for inferring the system trajectories $\bm{x}(t)$ and parameters $\bm{\theta}$ given the observed data $\bm{y}(\bm{\tau})$. The interested reader may refer to \citet{yang2021inference} for additional details.

The \pkg{MAGI} method places a GP prior on $\bm{x}(t)$, so that $\dot{\bm{x}}(t)$ conditional on $\bm{x}(t)$ also has a convenient GP form for facilitating inference without the need for numerical integration.  Previous authors adopting this basic idea, e.g.,  \citet{calderhead2009accelerating,dondelinger2013ode,barber2014gaussian,pmlr-v89-wenk19a}, have noted that this setup can cause $\dot{\bm{x}}(t)$ to be conceptually specified in two incompatible ways: first via the function $\mathbf{f}$ in the ODE (see Equation~\ref{eq:ode}), and second via the GP, e.g., as seen above in the setup of \pkg{deGradInfer}. The \pkg{MAGI} method addressed this conceptual difficulty by conditioning the GP on a manifold constraint that satisfies the ODEs specified by $\mathbf{f}$.

This manifold constraint can be described as follows. First, let $D$ denote the number of system components, with $x_d(t)$ and $\mathbf{f}(\bm{x}(t), \bm{\theta}, t)_d$ denoting the $d$-th component of $\bm{x}(t)$ and $\mathbf{f}$ respectively, $d=1, \ldots, D$, and let $C^1[0,T]$ be the set of differentiable functions on $[0,T]$. Then for $\bm{x}: [0,T] \to \mathbb{R}^D$ and a given parameter space $\Omega_{\bm{\theta}}$, we define a manifold $\mathcal{X}$ on which the derivative $\dot{\bm{x}}$ satisfies the dynamics specified by the ODE:
\begin{align*}
\mathcal{X} = \{ \bm{x} = (x_1, \ldots, x_D) \mid & \,  x_d \in C^1[0,T], \mbox{ for all } d = 1, \ldots, D,\\
& \dot{\bm{x}}(t) = \mathbf{f}(\bm{x}(t), \bm{\theta}, t) \, \mbox{ for all } t \in [0, T] \mbox{ and some } \bm{\theta} \in \Omega_{\bm{\theta}} \},
\end{align*}
i.e., $\bm{x} \in \mathcal{X}$ lies on the manifold of the ODE solutions. Second, to incorporate this manifold as a constraint in the GP, we define a variable $W$ according to
\begin{equation*}\label{eq:w}
W = \sup_{t \in [0,T], d \in \{1,\ldots, D\}} |\dot X_d(t) - \mathbf{f}(\bm{X}(t), \bm{\theta}, t)_d|,
\end{equation*}
i.e., $W$ quantifies the maximum discrepancy between a derivative trajectory and the dynamics implied by the ODEs. Thus, $W=0$ if and only if a realization $\bm{X}=\bm{x}$ of the GP satisfies $\bm{x} \in \mathcal{X}$. Under a Bayesian paradigm, the joint posterior distribution of $\bm{\theta}$ and $\bm{X}(t)$ is then conditioned on $W=0$ and the observed data $\bm{y}(\bm \tau)$, i.e., the ideal posterior of interest is $p(\bm{\theta}, \bm{x}(t) |  W = 0, \bm{y}(\bm \tau))$. However to be practically computable, $W=0$ needs to be approximated by finite discretization. Let $\bm{I} = \{ t_1, t_2, \ldots, t_n\}$ be a set of discretization points in  $[0,T]$, with $\bm{\tau} \subset \bm{I}$; then as a discretized analogue to $W$, we define a variable $W_{\bm{I}}$ according to
\begin{equation*}
W_{\bm{I}} = \max_{t \in \bm{I}, d \in \{1,\ldots, D\}} |\dot X_d(t) - \mathbf{f}(\bm{X}(t), \bm{\theta}, t)_d|.
\end{equation*}
This allows us to approximate $W=0$ by setting $W_{\bm{I}}=0$, i.e., $\dot{\bm{X}}(t)$ from the GP is constrained to equal $\mathbf{f}(\bm{X}(t), \bm{\theta}, t)$ from the ODE for each component $d=1, \ldots, D$ at each time point $t \in \bm{I}$. With $W_{\bm{I}}=0$ as the manifold constraint in practice, the corresponding posterior distribution is $p(\bm{\theta}, \bm{x}(\bm{I}) |  W_{\bm{I}} = 0, \bm{y}(\bm \tau))$. As we shall see below, this construction induces a closed-form posterior such that standard techniques of Bayesian inference can be applied, while ensuring that posterior samples of $\bm{x}(\bm{I})$ respect the ODE dynamics. Our construction also contrasts with penalized likelihood or regularization-based approaches \citep[e.g.,][]{ramsay2007parameter,pCODE_Rpack,niu2021r}; applying the rules of conditional probability with $W_{\bm{I}} = 0$ ensures that the ODE dynamics are exactly followed for all time points in $\bm{I}$, without the need to construct any penalty or regularization term.

\begin{figure}[t!]
	\centering
	\subcaptionbox{GP given observations only.
		\label{fig:GP-given-obs}}%
	[.495\linewidth]{\includegraphics[width=\linewidth]{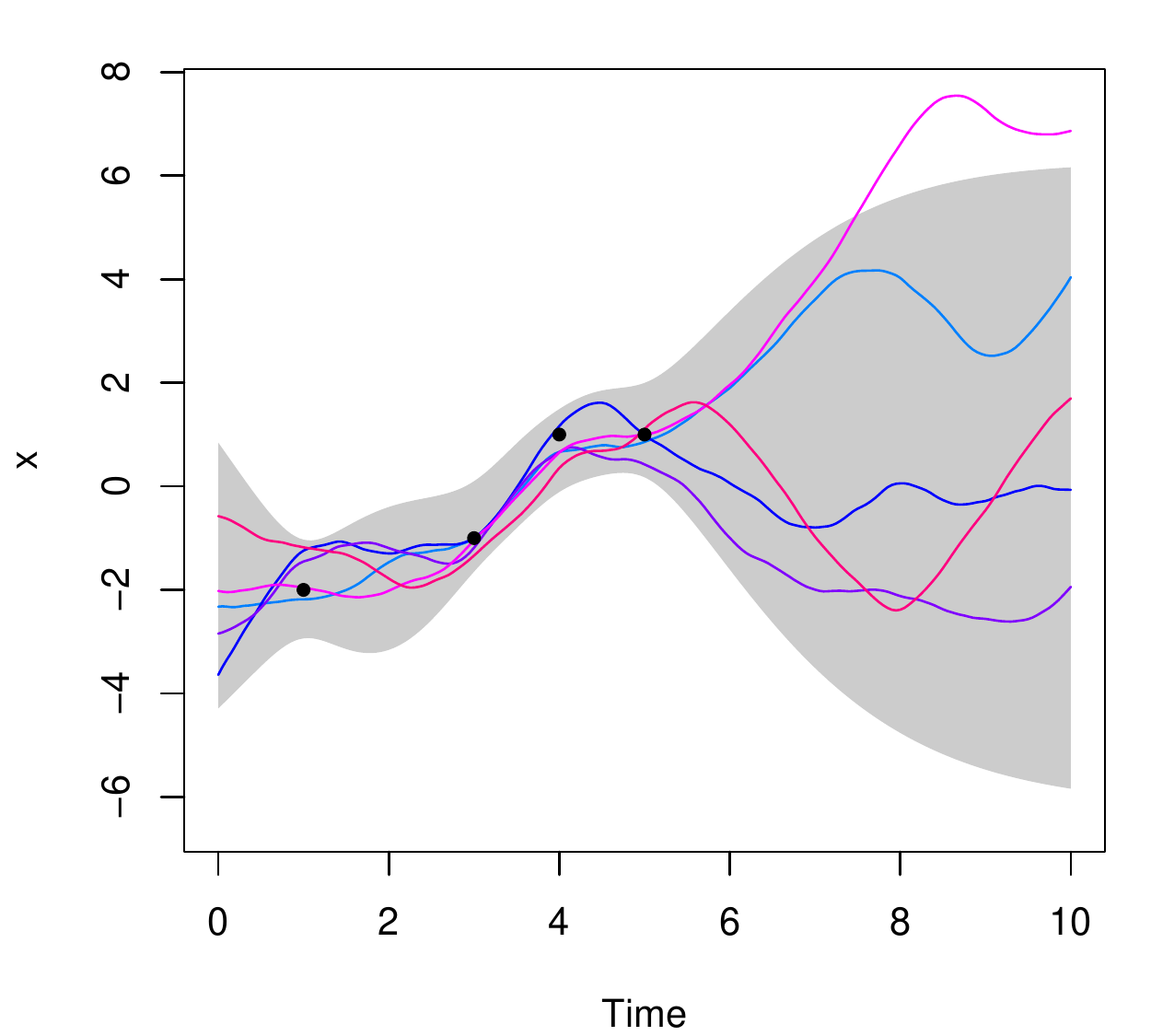}}
	\hfill
	\subcaptionbox{GP given observations and manifold constraint. \label{fig:GP-given-obs-constraint}}%
	[.495\linewidth]{\includegraphics[width=\linewidth]{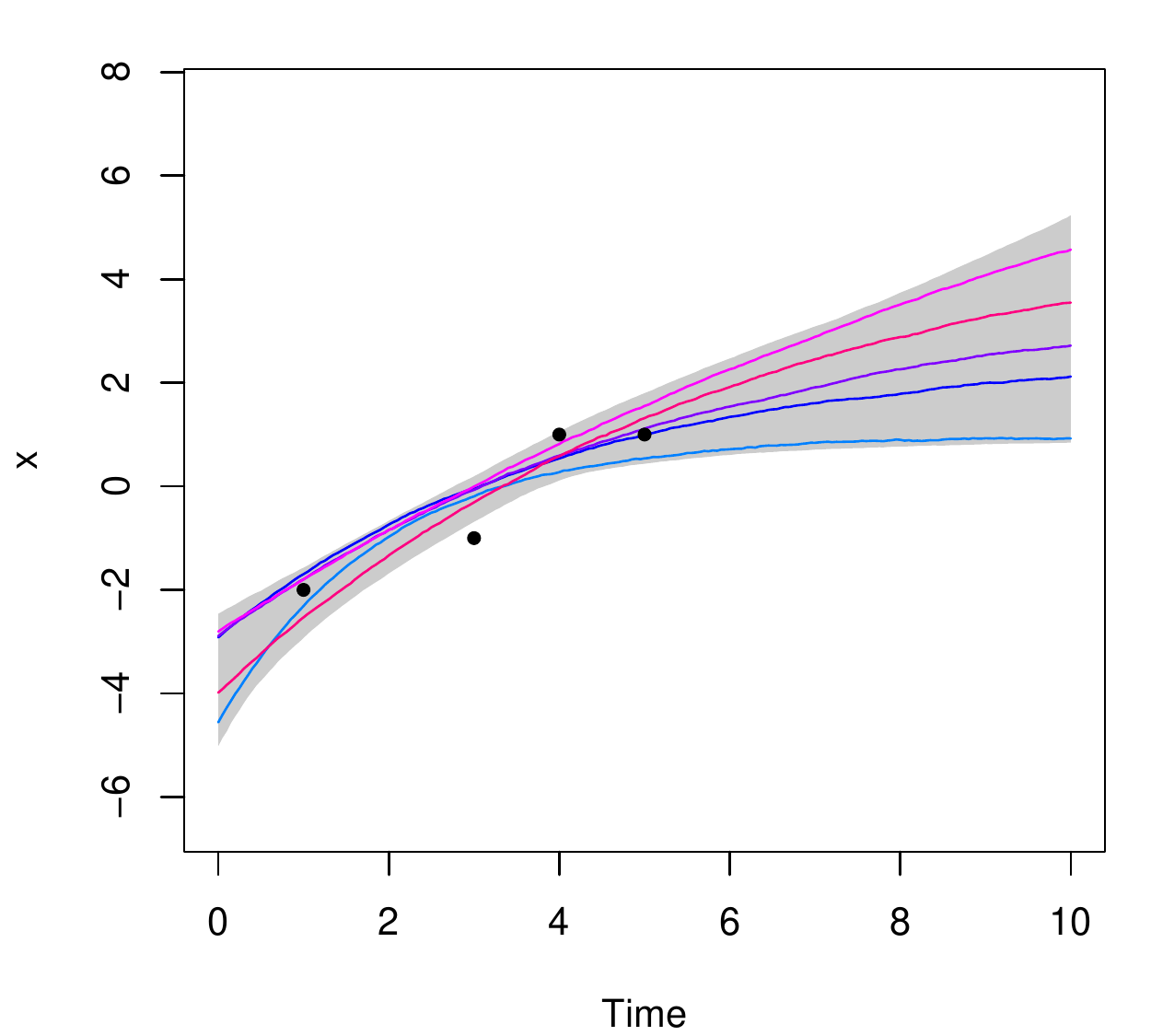}}
	\caption{Example visualization of the ODE manifold constraint. In panel (a), the GP is conditioned on four noisy observations (solid black dots). The colored curves are five random trajectories drawn from the resulting GP posterior, and the grey shaded area represents the 95\% credible interval. In panel (b), the GP is conditioned on the same four noisy observations and also the ODE manifold constraint. The colored curves are five random trajectories drawn from the resulting GP posterior, and the grey shaded area represents the 95\% credible interval with the manifold constraint.}
	\label{fig:GP-illustration}
\end{figure}

To illustrate this concept of manifold constraint, we guide the reader through a simple example in Figure \ref{fig:GP-illustration}. We begin with a selected GP prior on a 1-dimensional $x(t)$ over the interval $[0, 10]$. (For practical computation, the GP will be evaluated at $\bm{I} = \{ 0, 0.05, 0.1, \ldots, 10\}$.) Suppose four noisy observations $y(\bm{\tau})$ are taken at $\bm{\tau}=\{1, 3, 4, 5\}$; these are shown as black points in the panels of Figure \ref{fig:GP-illustration}. Then the GP posterior conditional on these four observations is visualized in Figure \ref{fig:GP-given-obs}. Five sample trajectories drawn from this GP posterior are shown via the colored curves, and the grey bands in Figure \ref{fig:GP-given-obs} represent 95\% credible intervals of this GP posterior. Next, as an example suppose $x(t)$ satisfies the linear ODE $ \dot{x}(t) = f({x}(t),\bm{\theta}, t)  = \theta_1 x(t) + \theta_2$ with parameters $\bm{\theta} = (\theta_1, \theta_2)$, which has analytic solution $x(t) = c \exp(\theta_1 t) - \theta_2 / \theta_1$ for some constant $c$. We now proceed to condition the GP on both $y(\bm{\tau})$ and the ODE manifold constraint $W_{\bm{I}}=0$. Five draws of $x(\bm{I})$ and $\bm{\theta}$ from this manifold-constrained GP posterior are visualized via the colored curves in Figure \ref{fig:GP-given-obs-constraint}. We see that they each obey the functional form of the known analytic solution $x(t)$ in this case, namely $x(t) = c \exp(\theta_1 t) - \theta_2 / \theta_1$ with different values of $c$, $\theta_1$, and $\theta_2$, i.e., $\dot{x}(t)$ from the GP satisfies the ODE specified by $f$. The grey shaded area in Figure \ref{fig:GP-given-obs-constraint} represents 95\% credible intervals of this manifold-constrained GP posterior. Contrasting the two panels of Figure \ref{fig:GP-illustration}, this analytical example provides an intuitive depiction for the key idea of the ODE manifold constraint. The \proglang{R} code to produce Figure \ref{fig:GP-illustration} is given in the replication script.

We return to our discussion of the joint posterior distribution on $\bm{x}(\bm{I})$  (i.e., the system trajectories at $t_1,\ldots, t_n$) and  $\bm{\theta}$, given $\bm{y}(\bm{\tau})$ and $W_{\bm{I}}=0$, which is expressed using Bayes' rule and factorized according to 
\begin{align*}
& p(\bm{\theta}, \bm{x}(\bm{I}) |  W_{\bm{I}} = 0, \bm{y}(\bm \tau)) \nonumber \\
& ~~~ \propto p({\bm{\Theta}} = \bm{\theta}, \bm{X}(\bm{I}) = \bm{x}(\bm{I}), W_{\bm{I}}=0, \bm{Y}(\bm\tau) = \bm{y}(\bm \tau)) \nonumber \\
& ~~~ = \pi(\bm{\theta}) \; \times \; p(\bm{X}(\bm{I}) = \bm{x({I})}  | {\bm{\Theta}} = \bm{\theta} ) \;
\times \; p(\bm{Y}(\bm\tau) = \bm{y}(\bm\tau) | \bm{X}(\bm I) = \bm{x}(\bm I), {\bm{\Theta}} = \bm{\theta}) \\
&\qquad \times p(W_{\bm{I}}=0 | \bm{Y}(\bm\tau) = \bm{y}(\bm\tau), \bm{X}(\bm I) = \bm{x}(\bm I), {\bm{\Theta}} = \bm{\theta}).
\end{align*}
Next, we note that both the GP prior for $\bm{X}$ and the observations are independent of $\bm{\Theta}$, so we have the simplifications $p(\bm{X}(\bm{I}) = \bm{x}(\bm{I}) | {\bm{\Theta}} = \bm{\theta} ) = p(\bm{X}(\bm{I}) = \bm{x}(\bm{I}) )$ and $p(\bm{Y}(\bm\tau) = \bm{y}(\bm\tau) | \bm{X}(\bm I) = \bm{x}(\bm I), {\bm{\Theta}} = \bm{\theta})  = p(\bm{y}(\bm\tau) | \bm{x}(\bm I))$. To simplify the last term, we substitute the definition of $W_{\bm{I}}=0$ and use the fact that the GP derivative $\bm{\dot{X}}(\bm{I})$ given $\bm{X}(\bm I)$ has a multivariate normal distribution that is conditionally independent of $\bm{\Theta}$ and the observations:
\begin{align*}
&  p(W_{\bm{I}}=0 | \bm{Y}(\bm\tau) = \bm{y}(\bm\tau), \bm{X}(\bm I) = \bm{x}(\bm I), {\bm{\Theta}} = \bm{\theta}) \\
&= p(\bm{\dot X}(\bm{I}) = \mathbf{f}(\bm{x}(\bm{I}), \bm{\theta}, \bm{I}) | \bm{Y}(\bm\tau) = \bm{y}(\bm\tau), \bm{X}(\bm I) = \bm{x}(\bm I), {\bm{\Theta}} = \bm{\theta} ) \\
&=  p(\bm{\dot X}(\bm{I}) = \mathbf{f}(\bm{x}(\bm{I}), \bm{\theta}, \bm{I}) | \bm{x}(\bm{I}) ).
\end{align*}
Thus, we finally obtain
\begin{align}
& p(\bm{\theta}, \bm{x}(\bm{I}) |  W_{\bm{I}} = 0, \bm{y}(\bm \tau)) \nonumber \\
& ~~~ \propto  \pi(\bm{\theta}) \times p( \bm{x}(\bm{I}) ) \times  p(  \bm{y}(\bm \tau) | \bm{x}(\bm{I})) \times p(\bm{\dot X}(\bm{I}) = \mathbf{f}(\bm{x}(\bm{I}), \bm{\theta}, t_{\bm{I}}) | \bm{x}(\bm{I}) ), \label{eq:main}
\end{align}
where the four terms are: $\pi(\bm{\theta})$ the prior density of the parameters, $p( \bm{x}(\bm{I}) )$ the multivariate normal density for the GP prior on $\bm{x}(t)$ evaluated at the points in  $\bm{I}$, $ p(  \bm{y}(\bm \tau) | \bm{x}(\bm{I}))$ the likelihood of the noisy observations, and $p(\bm{\dot X}(\bm{I}) = \mathbf{f}(\bm{x}(\bm{I}), \bm{\theta}, t_{\bm{I}}) | \bm{x}(\bm{I}) )$ the multivariate normal density for the conditional distribution of $\bm{\dot X}(\bm{I})$ given $\bm{X}(\bm{I})$ evaluated at $\bm{\dot X}(\bm{I}) = \mathbf{f}(\bm{x}(\bm{I}), \bm{\theta}, t_{\bm{I}})$. Equation~\ref{eq:main} is the basis of inference for \pkg{MAGI}. 

Note that the GP prior on $\bm{x}(t)$ involves hyper-parameters that govern the mean and covariance functions of the GP; we denote these hyper-parameters by $\bm{\phi}$. The likelihood $p(\bm{y}(\bm \tau) | \bm{x}(\bm{I}))$ may additionally depend on noise parameters, we denote these by $\bm{\sigma}$, which may be known or unknown depending on the application.

The \pkg{MAGI} method obtains Markov chain Monte Carlo (MCMC) samples for $\bm{x}(\bm{I})$ and $\bm{\theta}$ from Equation~\ref{eq:main}. The method proceeds according to the following steps:
\begin{enumerate}
	\item For system components that have observations, obtain values of $\bm{\phi}$ and $\bm{\sigma}$ by finding the optimal hyper-parameters and noise level based on fitting a GP to the data.   (If the noise level $\bm{\sigma}$ is known or given, optimization is applied to $\bm{\phi}$ only.)  The $\bm{\sigma}$ obtained is used to initialize MCMC sampling when the noise level is unknown.  MCMC sampling for $\bm{x}(\bm{I})$ is initialized by linearly interpolating the observations $\bm{y}(\bm{\tau})$.
	\item For system components that are unobserved, obtain values of $\bm{\phi}$ and $\bm{x}(\bm{I})$ together with $\bm{\theta}$ by a joint optimization of Equation~\ref{eq:main}, treating $(\bm{\phi},\bm{\sigma}, \bm{x}(\bm{I}))$ for the observed components obtained in step 1 as fixed.  If there are no unobserved components, only $\bm{\theta}$ is optimized in Equation~\ref{eq:main}.  This step  provides values at which  MCMC sampling for $\bm{\theta}$ is initialized, along with $(\bm{\phi},\bm{x}(\bm{I}))$ for unobserved components.
	\item Hamiltonian Monte Carlo \citep[HMC,][]{neal2011mcmc} is used as the MCMC sampling algorithm for obtaining joint draws of $\bm{x}(\bm{I})$ and $\bm{\theta}$ (together with $\bm{\sigma}$ if the noise parameters are unknown). During this posterior sampling, the GP hyper-parameters $\bm{\phi}$ are fixed at the values obtained in steps 1 and 2.  Our implementation of HMC automatically tunes the leapfrog step sizes of HMC during burn-in to achieve an acceptance rate of 60--90\%. (The theoretical optimal acceptance rate for HMC is 65\%, as suggested in \citet{neal2011mcmc}.)
\end{enumerate}

At the conclusion of MCMC sampling (and after discarding the burn-in iterations), we may treat the posterior means of $\bm{x}(\bm{I})$ and $\bm{\theta}$ as estimates of the trajectories and parameters respectively. The MCMC samples can also be used to characterize the uncertainty in these estimates, by computing credible intervals (CIs).

More methodological details of the \pkg{MAGI} method are discussed in \citet{yang2021inference}; the remainder of this paper focuses on practical usage of \pkg{MAGI} and the functionalities of the software package.

\section[Using the MAGI package]{Using the \pkg{MAGI} package} \label{sec:basicusage}

This section illustrates the main functionalities of the \pkg{MAGI} package by analyzing the sample dataset for the Hes1 model discussed in the Introduction.

After installing the \pkg{MAGI} package \citep{magi_Rpack} from CRAN, we load it into \proglang{R}
\begin{Sinput}
R> library("magi")
\end{Sinput}

The overall method is carried out via the core function \code{MagiSolver}, which initializes the GP hyper-parameters $\bm{\phi}$ and then carries out MCMC sampling for the parameters together with the system trajectories. The basic syntax is
\begin{Sinput}
R> MagiSolver(y, odeModel, control = list())
\end{Sinput}
where \code{y} is a data matrix that includes a column named \code{time} for the time points, \code{odeModel} is a list that specifies the ODE functions and its parameters, and \code{control} is a list used to provide any additional control settings. We describe each of these in turn, as we set up \code{MagiSolver} for the Hes1 dataset.

Since the Hes1 observations for $P$ and $M$ are positive and subject to  multiplicative error, we may apply a log-transform to the data and equations for the analysis, which then satisfies the framework for additive noise $\bm{\epsilon}$. Thus, continuing the data setup from Section \ref{sec:intro}, we create \code{y.tilde} as the log-transformed version of the observations \code{y} (i.e., excluding the \code{time} column):
\begin{Sinput}
R> y.tilde <- y
R> y.tilde[, names(y.tilde) != "time"] <-
      log(y.tilde[, names(y.tilde) != "time"])
\end{Sinput}
Recall that any unobserved values of \code{y.tilde} (in this case, the entire $H$ component column and every other value for $P$ and $M$) are assigned \code{NaN}.  The data matrix \code{y.tilde} for input to \code{MagiSolver} is now prepared, where the discretization set is $\bm{I} = \left\{ 0, 7.5, 15, \ldots, 240 \right\}$ minutes and corresponds to the time points where either $P$ or $M$ are observed. Note that with $|\bm{I}|$ denoting the cardinality of $\bm{I}$, the dimensions of the input data matrix are $|\bm{I}| \times (D+1)$, to include a column for time and each system component. To set up the input data matrix with different choices of $\bm{I}$, a helper function \code{setDiscretization} is provided; see Section \ref{subsec:discret} for a discussion of its usage and general guidelines for setting up $\bm{I}$ in practice. 

Next we set up the functions required for the \code{odeModel} list.  First, we apply a log-transform to each component of the ODE system by defining $\tilde X = (\log(P), \log(M), \log(H))$, then applying the relation $ \frac{d (\log u)}{dt} = \frac{du}{dt} \cdot \frac{1}{u}$ we see that $\tilde X$ satisfies 
\begin{align}
\mathbf{f}(\tilde X, \bm{\theta}, t) = \begin{pmatrix}
-aH + bM/P - c \\
-d + \frac{e}{(1 + P^2)M} \\
-aP + \frac{f}{(1+ P^2)H} - g
\end{pmatrix}, \label{eq:hes1log}
\end{align}
where $P$, $M$, and $H$ are the components in Equation~\ref{eq:hes1nolog}. We may code this via the function
\begin{Sinput}
R> hes1logmodelODE <- function (theta, x, tvec) {
+    P = exp(x[, 1])
+    M = exp(x[, 2])
+    H = exp(x[, 3])
+ 
+    PMHdt <- array(0, c(nrow(x), ncol(x)))
+    PMHdt[, 1] = -theta[1] * H + theta[2] * M / P - theta[3]
+    PMHdt[, 2] = -theta[4] + theta[5] / (1 + P^2) / M
+    PMHdt[, 3] = -theta[1] * P + theta[6] / (1 + P^2) / H - theta[7]
+ 
+    PMHdt
+  }
\end{Sinput}
which follows the same format as the function for the original (non-transformed) \code{hes1modelODE}.

Second, to facilitate MCMC sampling via HMC we also supply functions for the gradients of the ODEs with respect to the system components \code{x} and the parameters \code{theta}.  With respect to \code{x}, we have the matrix of gradients as follows,
\begin{align*}
\frac{\partial \mathbf{f}(\tilde X, \bm{\theta}, t)}{\partial \tilde  X} = \begin{pmatrix}
-b M/P & b M/P & -a H \\
-\frac{2eP^2}{ ( 1 + P^2)^2 M}  & -\frac{e}{(1 + P^2)M} & 0 \\
-aP - \frac{2fP^2}{ ( 1 + P^2)^2 H} & 0 &  -\frac{f}{(1 + P^2)H} \\
\end{pmatrix},
\end{align*}
which are specified via a function that outputs a 3-D array with dimensions $ |\bm{I}| \times D \times D$, where the array slice \code{[, i, j]} is the partial derivative of the ODE for the $j$-th system component with respect to the $i$-th system component:

\begin{Sinput}
R> hes1logmodelDx <- function (theta, x, tvec) {
+    logP = x[, 1]
+    logM = x[, 2]
+    logH = x[, 3]
+   
+    Dx <- array(0, c(nrow(x), ncol(x), ncol(x)))
+    dP = -(1 + exp(2 * logP))^(-2) * exp(2 * logP) * 2
+    Dx[, 1, 1] = -theta[2] * exp(logM - logP)
+    Dx[, 2, 1] = theta[2] * exp(logM - logP)
+    Dx[, 3, 1] = -theta[1] * exp(logH)
+    Dx[, 1, 2] = theta[5] * exp(-logM) * dP
+    Dx[, 2, 2] = -theta[5] * exp(-logM) / (1 + exp(2 * logP))
+    Dx[, 1, 3] = -theta[1] * exp(logP) + theta[6] * exp(-logH) * dP
+    Dx[, 3, 3] = -theta[6] * exp(-logH) / (1 + exp(2 * logP))
+   	
+    Dx
+  }
\end{Sinput}

With respect to \code{theta}, we have the matrix of gradients as follows, 
\begin{align*}
\frac{\partial \mathbf{f}( \tilde X, \bm{\theta}, t)}{\partial\bm{\theta}} = \begin{pmatrix}
-H &  M/P & -1 &  0 & 0 & 0  & 0 \\
0 & 0 & 0 & -1 &  \frac{1}{ ( 1 + P^2)^2 M} & 0 & 0 \\
-P & 0 & 0 & 0 & 0 & \frac{1}{ ( 1 + P^2)^2 H} & -1 
\end{pmatrix},
\end{align*}
which are specified via a function that outputs a 3-D array with dimensions $ |\bm{I}| \times |\bm{\theta}| \times D$, where the array slice \code{[, i, j]} is the partial derivative of the ODE for the $j$-th system component with respect to the $i$-th parameter in $\bm{\theta}$:
\begin{Sinput}
R> hes1logmodelDtheta <- function (theta, x, tvec) {
+    logP = x[, 1]
+    logM = x[, 2]
+    logH = x[, 3]
+   
+    Dtheta <- array(0, c(nrow(x), length(theta), ncol(x)))
+    Dtheta[, 1, 1] = -exp(logH)
+    Dtheta[, 2, 1] = exp(logM - logP)
+    Dtheta[, 3, 1] = -1
+    Dtheta[, 4, 2] = -1
+    Dtheta[, 5, 2] = exp(-logM) / (1 + exp(2 * logP))
+    Dtheta[, 1, 3] = -exp(logP)
+    Dtheta[, 6, 3] = exp(-logH) / (1 + exp(2 * logP))
+    Dtheta[, 7, 3] = -1
+   	
+    Dtheta
+  }
\end{Sinput}
At this point, it can be worthwhile to check that the gradients have been coded correctly. This can be done using \code{testDynamicalModel}, which tests the provided analytic gradients for correctness using numerical differentiation (via a finite difference approximation). To illustrate, we generate some test values for the data and \code{theta} as input into \code{testDynamicalModel} along with our ODE functions,
\begin{Sinput}
R> yTest <- matrix(runif(nrow(y.tilde) * (ncol(y.tilde) - 1)),
+                 nrow = nrow(y.tilde), ncol = ncol(y.tilde) - 1)
R> thetaTest <- runif(7)
R> testDynamicalModel(hes1logmodelODE, hes1logmodelDx, hes1logmodelDtheta, 
+                    "Hes1 log", yTest, thetaTest, y.tilde[, "time"])
\end{Sinput}
\begin{Soutput}
Hes1 log model, with derivatives
Dx and Dtheta appear to be correct
\end{Soutput}

which indicate the numerical and analytic gradients match for both \code{hes1logmodelDx} and \code{hes1logmodelDtheta}.

Third, \code{odeModel} must specify the upper and lower bounds on the parameters \code{theta}.  In this example, all of the seven parameters $(a,b,c,d,e,f,g)$ are non-negative, so we may set the corresponding bounds as \code{0} and \code{Inf}. 

We are now ready to define the required  list containing the three ODE model functions and parameter bounds:
\begin{Sinput}
R> hes1logmodel <- list(
+    fOde = hes1logmodelODE,
+    fOdeDx = hes1logmodelDx,
+    fOdeDtheta = hes1logmodelDtheta,
+    thetaLowerBound = rep(0, 7),
+    thetaUpperBound = rep(Inf, 7)
+  )
\end{Sinput}

Finally, additional settings can be supplied to \code{MagiSolver} via the list \code{control}, which may include any number of the following optional inputs. Brief descriptions are provided here, along with references to subsequent sections for further details.
\begin{itemize}
	\item Settings related to the MCMC sampling setup and initialization of $\bm{\sigma}$, $\bm{\theta}$ and $\bm{x}(\bm{I})$.
\begin{itemize}
	\item \code{sigma}: a numeric vector of length $D$, specifies the noise levels $\bm{\sigma}$ (i.e., standard deviations of observation noise) at which to initialize MCMC sampling.  By default, \code{MagiSolver} assumes that $\bm{\sigma}$ is unknown and initializes it via fitting a GP to the data. If the noise levels are known, supply \code{sigma} together with the option \code{useFixedSigma = TRUE}, which will then omit $\bm{\sigma}$ from MCMC sampling.
	\item \code{useFixedSigma}: logical, set to \code{TRUE} if $\bm{\sigma}$ is known.  If \code{useFixedSigma = TRUE}, the known values of $\bm{\sigma}$ must be supplied via the \code{sigma} control setting. Default is \code{FALSE}.
	\item \code{xInit}: a numeric matrix with dimension $|\bm{I}|\times D$, specifies values for the system trajectories at which to initialize MCMC sampling. Default is linear interpolation between the observed (non-missing) values of \code{y} to match the resolution of the discretization set $\bm{I}$.  An optimization routine is applied to Equation~\ref{eq:main} (as a function of $\bm{\theta}$, $\bm{\phi}$ and $\bm{x}(\bm{I})$ for unobserved system components) to initialize any unobserved system components.	
	\item \code{theta}: a numeric vector of the same length as $\bm{\theta}$, specifies values for the parameters $\bm{\theta}$ at which to initialize MCMC sampling. By default, \code{MagiSolver} optimizes Equation~\ref{eq:main} as a function of $\bm{\theta}$ only (with \code{xInit} fixed) to initialize \code{theta}; if there are unobserved system components, \code{theta} is initialized together with them (see \code{xInit}).
	\item \code{priorTemperature}:  numeric, a tempering factor by which to scale the contribution of the GP prior, to control the influence of the GP prior relative to the likelihood of the observations. Effectively, the log of the GP prior is divided by \code{priorTemperature}. Default is the total number of observations divided by the total number of discretization points, aggregated over all components; a fuller discussion is provided in Section \ref{subsec:discret}.
\end{itemize}	
	\item Settings related to the GP prior and its hyper-parameters. These options are discussed in detail in Section~\ref{subsec:phi}.
	\begin{itemize}
	\item \code{kerneltype}: string, specifies the type of GP covariance function to use. The default and recommended choice (\code{generalMatern}) is a Matern kernel with degree of freedom 2.01; it has hyper-parameters $\phi_1$ and $\phi_2$ for each component that control the variance level and bandwidth, respectively. See Section~\ref{subsec:phi} for further discussion regarding this choice; other available choices for \code{kerneltype} are listed in Appendix~\ref{app:covariance}.
	\item \code{phi}: a numeric matrix with dimension $|\bm{\phi}_d| \times D$, specifies the values of the GP hyper-parameters $\bm{\phi}$, where $|\bm{\phi}_d|$ is the number of hyper-parameters for each component (i.e., $|\bm{\phi}_d|= 2$ for \code{generalMatern}).  By default, \code{MagiSolver} will estimate $\bm{\phi}$ automatically for observed components via GP fitting and for unobserved system components via optimization of Equation~\ref{eq:main} (see \code{xInit}).
	\item \code{mu}: a numeric matrix with dimension $|\bm{I}|\times D$, specifies values for the mean function of the GP prior of each component. Default is a zero mean function.  To use a custom mean function, \code{mu} must be specified together with \code{dotmu}.
	\item \code{dotmu}: a numeric matrix with dimension $|\bm{I}|\times D$, specifies values for the derivatives of the GP prior mean function for each component. Default is zero.
	\item \code{bandSize}:  integer, specifies the size of the diagonal band matrix approximation used to speed up matrix operations. Default \code{bandSize} is 20, can be increased if \code{MagiSolver} returns an error indicating numerical instability.	
	\end{itemize}
	\item Settings related to the Hamiltonian Monte Carlo (HMC) sampling algorithm that is used to obtain MCMC draws from the posterior. A description of HMC and the role of these options is provided in Section~\ref{subsec:hmc}.
	\begin{itemize}
	\item \code{niterHmc}: integer, specifies the number of HMC iterations to run. Default is 20000.
	\item \code{nstepsHmc}: integer, specifies the number of leapfrog steps per HMC iteration. Default is 200.
	\item \code{burninRatio}: numeric, specifies the proportion of HMC iterations to be discarded as burn-in. Default is 0.5, which discards the first half of the MCMC samples.
	\item \code{stepSizeFactor}: numeric, initial leapfrog step size factor for HMC. Default is 0.01, and the leapfrog step size is automatically tuned during burn-in to achieve an acceptance rate between 60-90\%.

	\end{itemize}

	\item Other miscellaneous settings for specialized situations.
	\begin{itemize}
	\item \code{skipMissingComponentOptimization}: logical, set to \code{TRUE} to override automatic optimization for unobserved components.  If \code{skipMissingComponentOptimization = TRUE}, values for \code{xInit} and \code{phi} must be supplied for all system components. Default is \code{FALSE}.
	\item \code{positiveSystem}: logical, set to \code{TRUE} to enforce the constraint that  $\bm{x}(\bm{I})$ is non-negative for all system components. Default is \code{FALSE}.
	\item \code{verbose}: logical, set to \code{TRUE} to output diagnostic and sampling progress messages to the console. Default is \code{FALSE}.	

	\end{itemize}
\end{itemize}

For most settings, the defaults are generally recommended as a reasonable starting point for using \code{MagiSolver}. The examples provided in this paper will illustrate some cases where it is necessary to override the defaults.

In the Hes1 example, the noise standard deviations are known. We supply these values via \code{sigma} and set \code{useFixedSigma = TRUE} in the \code{control} list, as otherwise  $\bm{\sigma}$ is treated as a parameter that is sampled within each HMC iteration. We use the defaults for the remaining settings and run \code{MagiSolver} on the Hes1 dataset as follows, which stores the output in \code{hes1result}:
\begin{Sinput}
R> hes1result <- MagiSolver(y.tilde, hes1logmodel, 
+      control = list(sigma = param.true$sigma, useFixedSigma = TRUE))
\end{Sinput}

In \proglang{R}, the output of \code{MagiSolver} is an \proglang{S}3 object of class `\code{magioutput}' which contains the following list elements:
\begin{itemize}
	\item \code{theta}: matrix of MCMC samples for $\bm{\theta}$ after burn-in.
	\item \code{xsampled}: array of MCMC samples for the system trajectories $\bm{x} (\bm{I})$ after burn-in.
	\item \code{sigma}: matrix of MCMC samples for $\bm{\sigma}$ after burn-in.
	\item \code{lp}: vector of log-posterior values at each HMC iteration, after burn-in.	
	\item \code{phi}: matrix of estimated GP hyper-parameters $\bm{\phi}$.	
	\item \code{y, tvec, odeModel}: the data matrix, time vector, and \code{odeModel} specification from the inputs to \code{MagiSolver}.
\end{itemize}
For convenience in \proglang{R}, `\code{magioutput}' objects have the following associated methods to provide basic inferences from the MCMC samples:
\begin{itemize}
\item \code{print()}: displays a brief summary of the settings used for the \code{MagiSolver} run.
\item \code{summary()}: generates a table of parameter estimates and credible intervals.
\item \code{plot()}: visualizes the inferred trajectories and credible bands for each component, or generates diagnostic traceplots (i.e., plots of MCMC sampled values vs.~the number of iterations) for the parameters.
\end{itemize}

We have allowed the hyper-parameters $\bm{\phi}$ to be automatically estimated in this example (including for the unobserved $H$ component), which is often sufficient in our experience. Further guidelines for setting the GP prior and hyper-parameters are discussed in Section \ref{subsec:phi}.

\begin{figure}[ht]
	\centering
	\includegraphics[width=5.5in]{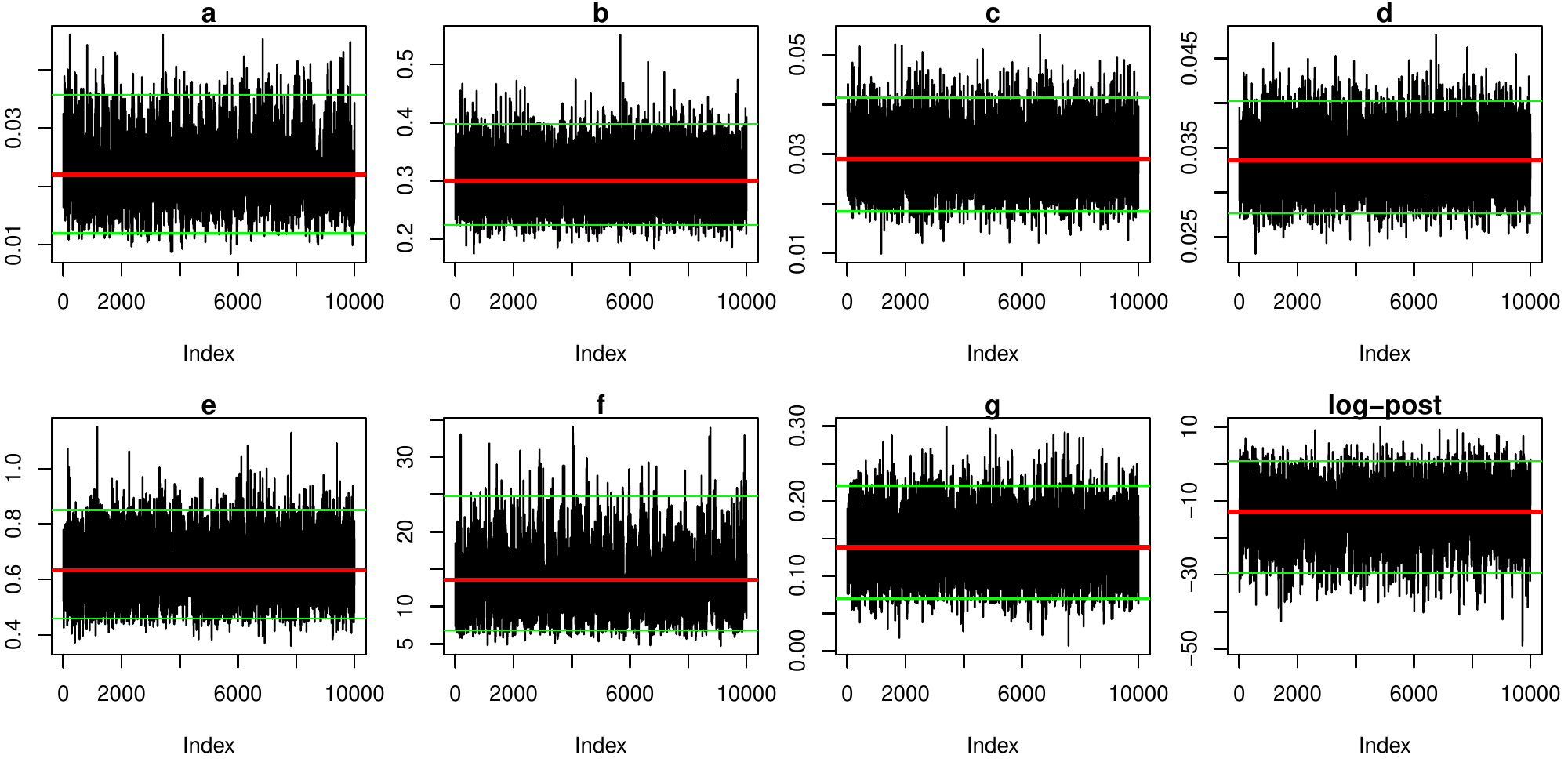}
	\caption{MCMC traceplots for the seven parameters of the Hes1 system and the log-posterior values. The horizontal lines in the plots indicate the posterior mean (red) and limits of the 95\% credible interval (green) for each parameter.} \label{fig:hes1trace}
\end{figure}

Turning to the MCMC samples, Figure \ref{fig:hes1trace} shows the traceplots of \code{theta} and \code{lp} (\code{sigma} is omitted since it is treated as known in this example) as an informal check for convergence, produced using the \code{plot()} convenience function with \code{type = "trace"}:
\begin{Sinput}
R> theta.names <- c("a", "b", "c", "d", "e", "f", "g")
R> plot(hes1result, type = "trace", par.names = theta.names, nplotcol = 4)
\end{Sinput}

The MCMC samples of each parameter randomly scatter around their posterior means (red horizontal lines), which visually indicate that convergence has occurred. The 95\% credible intervals (via 2.5 to 97.5 percentiles of the MCMC samples) are shown via the green horizontal lines. We generally suggest taking the posterior mean as the parameter estimate for $\bm{\theta}$.\footnote{Other commonly-used Bayesian point estimates include the median (which may be taken component-wise) and the mode (which may approximated by the MCMC sample with the highest log-posterior value). These can be obtained in \pkg{MAGI} by passing the \code{est} argument to the \code{plot()} or \code{summary()} methods. Since \pkg{MAGI} is based on a GP prior (and Gaussian tails are thin), the posterior mean (for both $\bm{\theta}$ and $\bm{x} (\bm{I})$) tends to be sufficiently robust and works well in practice.} The numerical values of these parameter estimates and credible intervals can be extracted using the convenience \code{summary()} method:
\begin{Sinput}
R> summary(hes1result, par.names = theta.names)
\end{Sinput}
\begin{Soutput}
           a     b      c      d     e     f      g
Mean  0.0220 0.300 0.0291 0.0336 0.634 13.50 0.1380
2.5%  0.0119 0.224 0.0185 0.0277 0.459  6.82 0.0694
97.5% 0.0358 0.398 0.0414 0.0403 0.851 24.80 0.2200
\end{Soutput}

The true parameter values are well contained in the 95\% credible intervals, with the exception of $g$, which only governs the unobserved $H$ component as seen in Equation~\ref{eq:hes1log}.

\begin{figure}[ht]
	\centering
	\includegraphics[width=\textwidth]{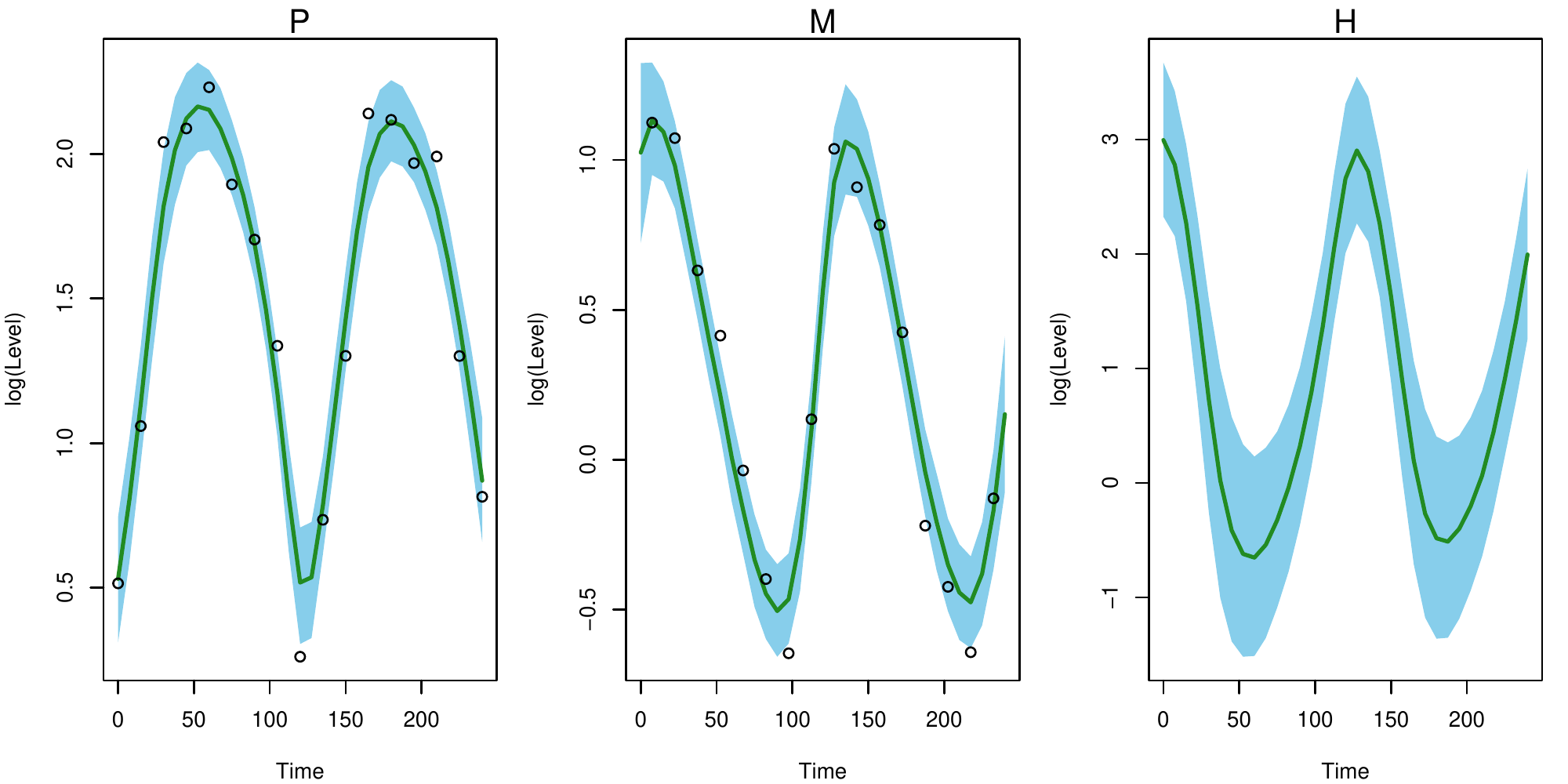}
	\caption{Inferred trajectories from \pkg{MAGI} for the three components of the Hes1 system on the log-scale (green curves), generated using the \code{plot()} method. The blue shaded areas represent 95\% credible intervals. The asynchronous noisy observations of $P$ and $M$ are plotted as black circles.} \label{fig:hes1defplot}
\end{figure}

Next, we can extract and visualize the sampled system trajectories $\bm{x} (\bm{I})$. We treat the posterior means as the inferred trajectories, and use the 2.5 to 97.5 percentiles of the MCMC samples to provide 95\% credible intervals at each time point in $\bm{I}$. These are the default settings of the convenience \code{plot()} method, which we use to generate Figure \ref{fig:hes1defplot}:
\begin{Sinput}
R> plot(hes1result, lwd = 2, col = "forestgreen", comp.names = compnames,
+         xlab = "Time", ylab = "log(Level)")
\end{Sinput}

\begin{figure}[ht]
	\centering
	\includegraphics[width=\textwidth]{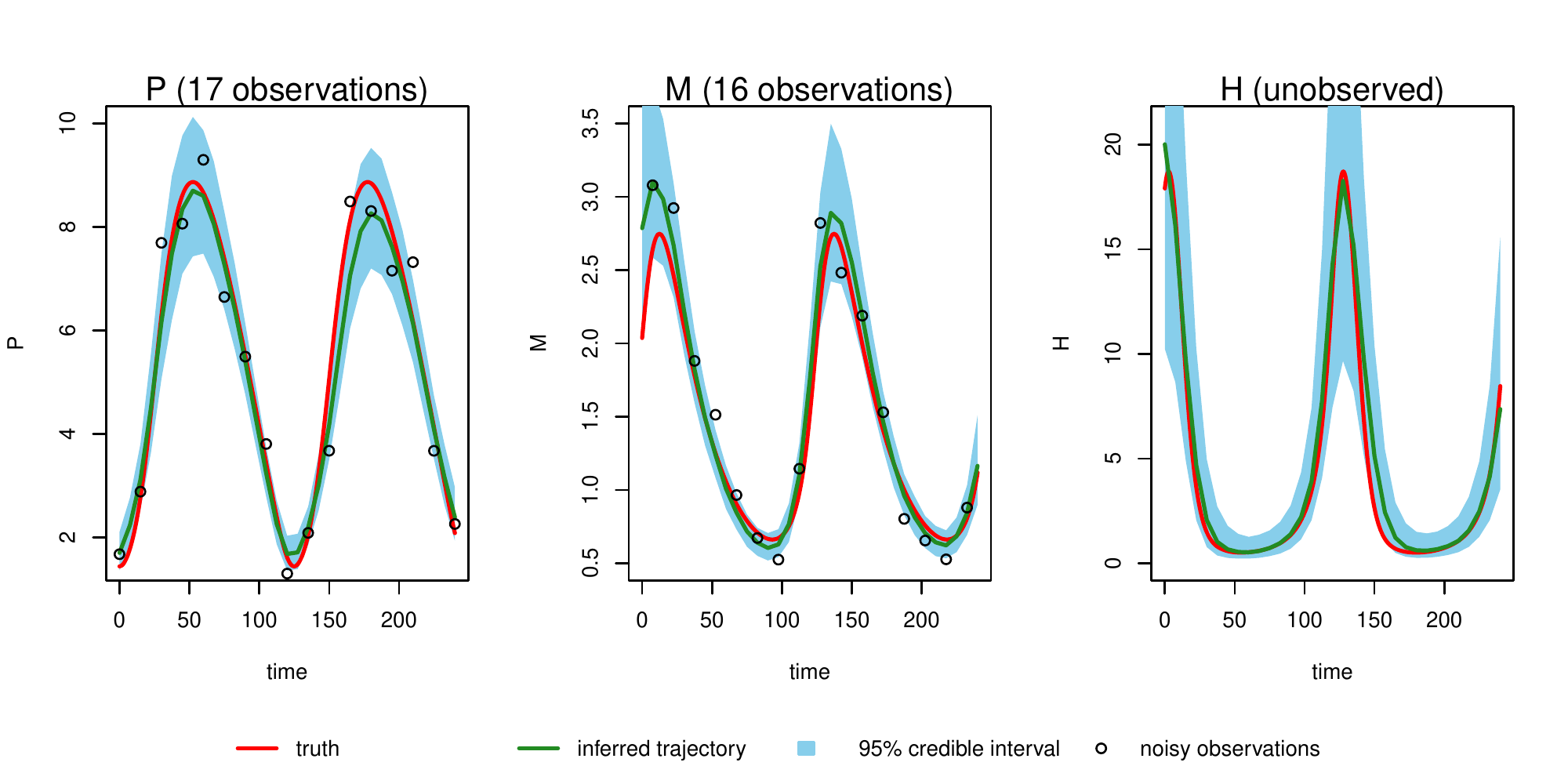}
	\caption{Inferred trajectories from \pkg{MAGI} for the three components of the Hes1 system (green curves). The blue shaded areas represent 95\% credible intervals. The asynchronous noisy observations of $P$ and $M$ are plotted as black circles, and the red curves represent the true underlying trajectories.} \label{fig:hes1traj}
\end{figure}

In this example it is helpful to do some further post-processing, and generate a customized plot on the original scale with the true trajectory overlaid. We exponentiate to convert each component to the original scale of measurement: 
\begin{Sinput}
R> xLB <- exp(apply(hes1result$xsampled, c(2, 3),
+                     function(x) quantile(x, 0.025)))
R> xMean <- exp(apply(hes1result$xsampled, c(2, 3), mean))
R> xUB <- exp(apply(hes1result$xsampled, c(2, 3),
+                     function(x) quantile(x, 0.975)))
\end{Sinput}
The inferred trajectories (green curves) and blue shaded areas representing the 95\% credible intervals are shown in Figure \ref{fig:hes1traj}, with the noisy observations and true trajectories overlaid as black points and red curves, respectively. The system trajectories are recovered well: the green curves for $P$ and $M$ are consistent with the observed data points and the truth for the entirely unobserved $H$ component is correctly inferred. The plotting code for Figure \ref{fig:hes1traj} is given in the replication script.

\section{Finer control of inference: features and examples}

This section presents two additional dynamic system examples to illustrate the role of the discretization set $\bm{I}$, the GP prior and its hyper-parameters $\bm{\phi}$, and the HMC algorithm. We discuss how to use these features to obtain finer control over the inference results. The second example (in Section \ref{subsec:phi}) also demonstrates a system with equations that explicitly depend on time.

\subsection{Choice of discretization set} \label{subsec:discret}

\pkg{MAGI} constrains the GP to satisfy the ODE system derivatives at the points in the discretization set $\bm{I}$. Therefore, increasing the denseness of $\bm{I}$ may lead to more accurate inference in some cases, with the trade-off being longer computation time. In practice, it can be a useful strategy to consider running \code{MagiSolver} with an increasing sequence of discretization sets  $\bm{I}_0 \subset \bm{I}_1 \subset \cdots $ to ensure that the estimates obtained are stable.

For the initial run, we recommend taking $\bm{I}_0$ as the smallest evenly-spaced set (or approximately so) that includes the observation time points $\bm{\tau}$. This can help ensure that the system dynamics are adequately captured throughout the modelled time interval --- note that this is not a requirement for running \code{MagiSolver} itself, which can handle any kind of spacing between time points. Then, we can construct subsequent sets $\bm{I}_j$, $j \ge 1$ by inserting one equally-spaced point between each pair of adjacent time points in $\bm{I}_{j-1}$.

The function \code{setDiscretization} can be used to prepare data matrices \code{y} according to this strategy. The command \code{setDiscretization(y, level = j)} returns a data matrix with  $2^j - 1$ equally-spaced points inserted between each observation of \code{y} (i.e., $j=0$ returns the original matrix \code{y}); this works well when \code{y} are evenly spaced. The alternative syntax \code{setDiscretization(y, by = incr)} can be useful when the observations in \code{y} are unevenly spaced: it returns a data matrix with time points inserted (as needed) to form an equally-spaced discretization set from the first to last observations of \code{y}, with interval \code{incr} between successive discretization points.

Mathematically, as the discretization set becomes more dense, the contributions of the terms in Equation~\ref{eq:main} associated with $\bm{I}$, i.e., the GP prior $p( \bm{x}(\bm{I}) )$ and $p(\bm{\dot X}(\bm{I}) = \mathbf{f}(\bm{x}(\bm{I}), \bm{\theta}, t_{\bm{I}}) | \bm{x}(\bm{I}) )$, would become larger relative to the likelihood of the observations. This is because the likelihood term in Equation~\ref{eq:main} simplifies as $p(\bm{y}(\bm \tau) | \bm{x}(\bm{I})) = p(  \bm{y}(\bm \tau) | \bm{x}(\bm{\tau}))$ for any $\bm{\tau} \subset \bm{I}$ and does not change as $\bm{I}$ becomes more dense, i.e., only the points in $\bm{I}$ corresponding to observations have an associated likelihood contribution. Therefore, \pkg{MAGI} automatically uses a tempering hyper-parameter $\beta$ to maintain the balance between the GP prior and the likelihood across different discretization sets. This helps ensure that parameter inference reaches a stable result over an increasing sequence of discretization sets. Specifically, $p( \bm{x}(\bm{I}) ) p(\bm{\dot X}(\bm{I}) = \mathbf{f}(\bm{x}(\bm{I}), \bm{\theta}, t_{\bm{I}}) | \bm{x}(\bm{I}) )$ is tempered as $ \left[p( \bm{x}(\bm{I}) ) p(\bm{\dot X}(\bm{I}) = \mathbf{f}(\bm{x}(\bm{I}), \bm{\theta}, t_{\bm{I}}) | \bm{x}(\bm{I}) ) \right]^{1/\beta}$, where our recommended value $\beta = D |\bm{I}| / \sum_{d=1}^D |\bm{\tau}_d |$ is the total number of discretization points divided by the total number of observations (aggregated over all components). For example if $|\bm{I}|$ is doubled, then tempering effectively reduces the GP contribution on the log-scale by half to compensate. (A custom value for $\beta$ can be set by the user via  \code{priorTemperature} in the optional \code{control} list to \code{MagiSolver}, but this is not generally recommended.)

We illustrate this idea of increasing discretization sets, on a dataset of noisy observations simulated from the classic FitzHugh-Nagumo (FN) equations for $X = (V, R)$ that model spike potentials of neurons \citep{fitzhugh1961impulses}:
\begin{align}
\mathbf{f}(X, \bm{\theta}, t)= \begin{pmatrix}
	c(V-\dfrac{V^3}{3}+R) \\
	-\dfrac{1}{c}(V-a+bR)
\end{pmatrix},\label{eq:FN}
\end{align}
where $V$ and $R$ are the voltage and recovery variables, and $\bm{\theta}= (a,b,c)$ are the parameters to be estimated.

We begin by loading the dataset and setting a random seed for reproducibility:
\begin{Sinput}
R> data("FNdat")
R> set.seed(12321)
\end{Sinput}
The observation time points of \code{FNdat} are $t = 0, 0.5, 1, \ldots, 10$ at intervals of 0.5, along with $t = 11, 12, 13, 14, 15, 17, 20$. Following the suggested strategy above, we create a data matrix corresponding to the first discretization set $\bm{I}_0$, taken to be the 41 equally-spaced points $\{ 0, 0.5, 1, \ldots, 20\}$:
\begin{Sinput}
R> y_I0 <- setDiscretization(FNdat, by = 0.5)
\end{Sinput}
We also create data matrices corresponding to the denser sets $\bm{I}_1, \bm{I}_2, \bm{I}_3$ by successively inserting one equally-spaced time point between existing ones:
\begin{Sinput}
R> y_I1 <- setDiscretization(y_I0, level = 1)
R> y_I2 <- setDiscretization(y_I0, level = 2)
R> y_I3 <- setDiscretization(y_I0, level = 3)
\end{Sinput}

The gradients of the ODEs with respect to $X$ and $\bm{\theta}$ are as follows:
$$
\frac{\partial \mathbf{f}(X, \bm{\theta}, t)}{\partial X} = \begin{pmatrix}
c(1-V^2) & c \\
-1/c & -b/c
\end{pmatrix}
$$

$$
\frac{\partial \mathbf{f}(X, \bm{\theta}, t)}{\partial \bm{\theta}} = \begin{pmatrix}
0 & 0 & V-V^3/3 + R \\
1/c & -R/c & (V - a + bR)/c^2
\end{pmatrix}
$$

Functions that code the FN equations (\code{fnmodelODE}) and their gradients (\code{fnmodelDx} and \code{fnmodelDtheta}) are set up analogously to the Hes1 model, and are shown in Appendix \ref{app:FN}. Using these, we create the \code{odeModel} list input:
\begin{Sinput}
R> fnmodel <- list(
+    fOde = fnmodelODE,
+    fOdeDx = fnmodelDx,
+    fOdeDtheta = fnmodelDtheta,
+    thetaLowerBound = c(0, 0, 0),
+    thetaUpperBound = c(Inf, Inf, Inf)
+  )
\end{Sinput}

We can now run \code{MagiSolver} with the discretization sets we constructed and 10000 HMC iterations. Since the noise level is unknown in this dataset, $\bm{\sigma}$ will also be inferred via MCMC sampling. Note that the dimensionality of the variables being sampled effectively doubles with each successive discretization set. Thus, the outputs of the HMC iterations can become increasingly `sticky' (i.e., having higher autocorrelation) with denser discretization sets. For more detail on this point, see Section \ref{subsec:hmc} for a discussion of HMC and its settings.  One way to circumvent this is to increase the number of leapfrog steps per HMC iteration. We have illustrated that below, by setting \code{nstepsHmc = 1000} for our densest set $\bm{I}_3$ (otherwise the default is 200 leapfrog steps).
\begin{Sinput}
R> FNres0 <- MagiSolver(y_I0, fnmodel, control=list(niterHmc = 10000))
R> FNres1 <- MagiSolver(y_I1, fnmodel, control=list(niterHmc = 10000))
R> FNres2 <- MagiSolver(y_I2, fnmodel, control=list(niterHmc = 10000))
R> FNres3 <- MagiSolver(y_I3, fnmodel,
+                control=list(niterHmc = 10000, nstepsHmc = 1000))
\end{Sinput}
To compare the estimates, we make use of \code{summary()} to extract the posterior means and 95\% credible intervals for both $\bm{\theta}$ and $\bm{\sigma}$ from each model fit:
\begin{Sinput}
R> FNpar.names <- c("a", "b", "c", "sigmaV", "sigmaR")
R> FNsummary <- lapply(list(FNres0, FNres1, FNres2, FNres3),
+               function(x) summary(x, sigma = TRUE, par.names = FNpar.names))
\end{Sinput}
We then plot these posterior summaries for each parameter and discretization set:
\begin{Sinput}
R> layout(rbind(c(1:5), rep(6, 5)), heights = c(5, 0.25))
R> for (i in 1:length(FNpar.names)) {
+   par(mar = c(2, 4, 1.5, 1))
+   estCI <- sapply(FNsummary, function(x) x[,i])
+   plot(1:4, xlim = c(0, 5), ylim = c(min(estCI[2, ]), max(estCI[3, ])),
+        xaxt = 'n', xlab = '', ylab = '', type = 'n')
+   segments(1:4, y0 = estCI[2, ], y1 = estCI[3, ], col = 1:4, lwd = 2)
+   mtext(FNpar.names[i])
+   points(1:4, estCI[1, ], col = 1:4, cex = 2)
+ } 
R> par(mar = rep(0, 4))
R> plot(1, type = 'n', xaxt = 'n', yaxt = 'n',
+      xlab = NA, ylab = NA, frame.plot = FALSE)
R> legend("center", c("I0", "I1", "I2", "I3"),
+        col = 1:4, lwd = 4, horiz = TRUE, bty = "n")
\end{Sinput}

\begin{figure}[ht]
	\centering
	\includegraphics[width=\textwidth]{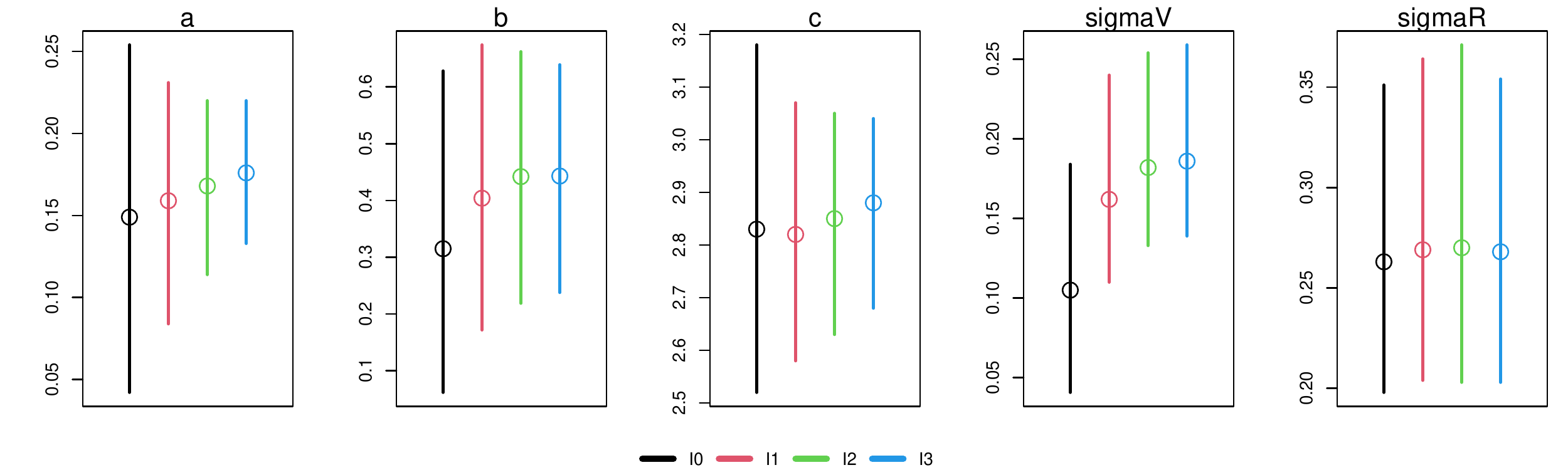}
	\caption{Posterior means (points) and 95\% credible intervals (vertical bars) for each parameter, from using \code{MagiSolver} on the simulated FN dataset with the discretization sets $\bm{I}_0 \subset \bm{I}_1 \subset \bm{I}_2 \subset \bm{I}_3$.} \label{fig:FNparam}
\end{figure}
The panels of Figure \ref{fig:FNparam} show that the posterior means (points) and credible intervals (vertical bars) visibly shift for $b$ and $\sigma_V$, as we use the successive discretization sets from $\bm{I}_0$ to  $\bm{I}_1$ to $\bm{I}_2$. Meanwhile comparing $\bm{I}_2$ and $\bm{I}_3$ for all the parameters, the posterior means are fairly similar and the credible intervals largely overlap, indicating that the inference results are stable.

\begin{figure}[ht]
	\centering
	\includegraphics[width=\textwidth]{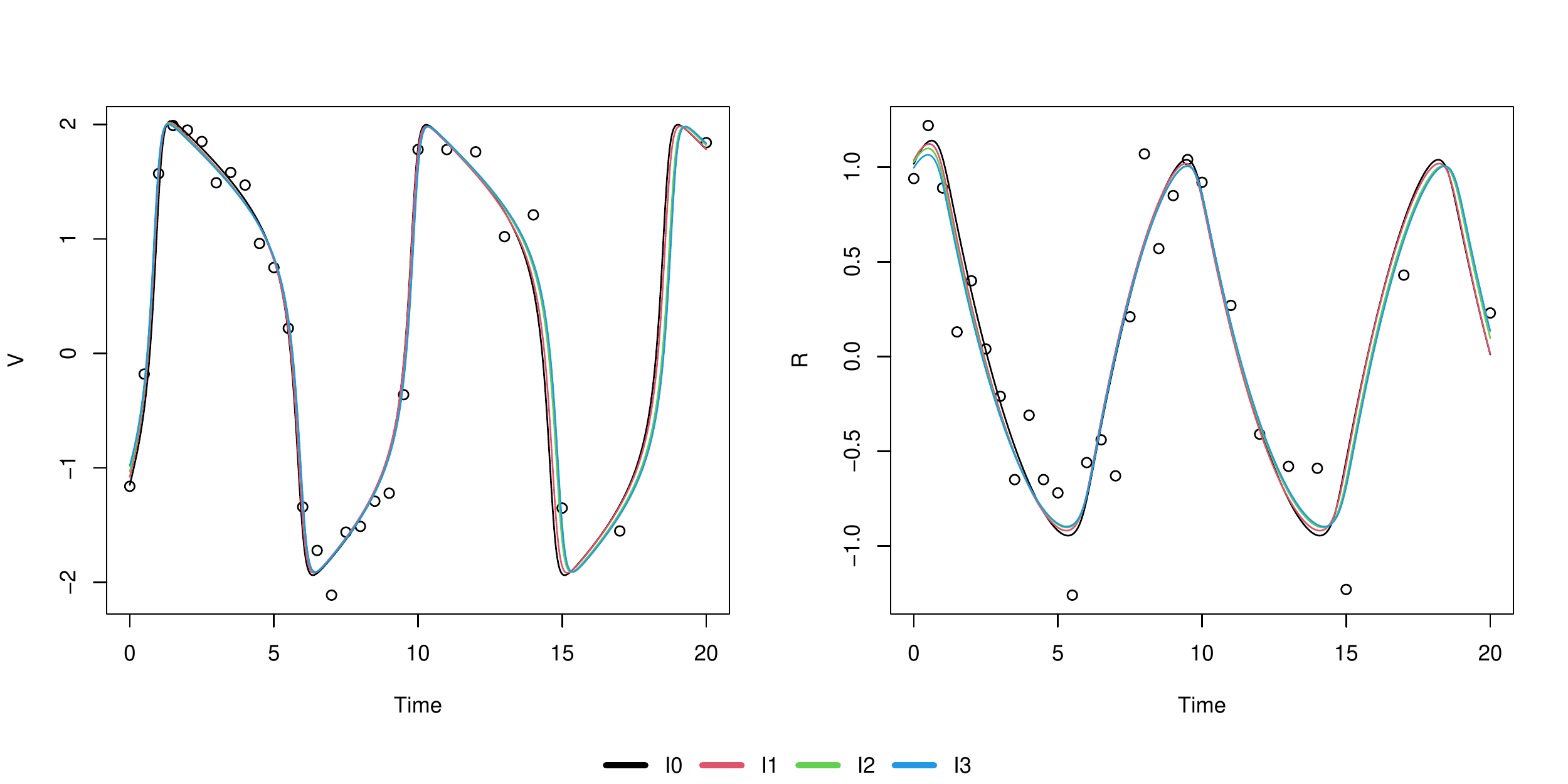}
	\caption{Reconstructed system trajectories (solid curves) based on the estimated parameters and initial conditions from the simulated FN dataset using \code{MagiSolver} with the discretization sets $\bm{I}_0 \subset \bm{I}_1 \subset \bm{I}_2 \subset \bm{I}_3$. The noisy observations (points) are also plotted. } \label{fig:FNtraj}
\end{figure}

To provide a further check, we can use the ODE solver to reconstruct the trajectories implied by the estimates of the parameters and the initial conditions. Again, we define a wrapper to facilitate calling \code{ode} from \pkg{deSolve}:
\begin{Sinput}
R> fnmodelODEsolve <- function(tvec, state, parameters) {
+    list(as.vector(fnmodelODE(parameters, t(state), tvec)))
+  }
\end{Sinput}
We define a helper function that invokes the ODE solver using the posterior means of $\bm{\theta}$ and $V(0), R(0)$ from the \code{MagiSolver} output. Note that the MCMC samples for the initial conditions can be extracted from the \code{xsampled} array, as shown below to obtain \code{x0.est}:
\begin{Sinput}
R> tvec <- seq(0, 20, by = 0.01)
R> FNcalcTraj <- function(res) {
+    x0.est <- apply(res$xsampled[, 1, ], 2, mean)
+    theta.est <- apply(res$theta, 2, mean)
+   
+    x <- deSolve::ode(y = x0.est, times = tvec,
+            func = fnmodelODEsolve, parms = theta.est)
+    x
+  }
\end{Sinput}
We then compute these reconstructed trajectories for the estimates based on the four discretization sets and plot them in Figure \ref{fig:FNtraj}. Visually, all four reconstructed trajectories fit the observed data well. There are some minor shifts in the trajectories as the discretization sets get denser from $\bm{I}_0$ to  $\bm{I}_1$ to $\bm{I}_2$, while those for $\bm{I}_2$ and $\bm{I}_3$ (green and blue) become nearly indistinguishable (except for $0 \le t \le 2$ of the $R$ component). The code to produce the figure is below:
\begin{Sinput}
R> FNtr <- lapply(list(FNres0, FNres1, FNres2, FNres3), FNcalcTraj)

R> layout(rbind(c(1, 2), c(3, 3)), heights = c(5, 0.25))
R> plot(FNdat$time, FNdat$V, xlab = "Time", ylab = "V")
R> matplot(tvec, sapply(FNtr, function(x) x[, 2]),
+    type = "l", lty = 1, add = TRUE)
R> plot(FNdat$time, FNdat$R, xlab = "Time", ylab = "R")
R> matplot(tvec, sapply(FNtr, function(x) x[, 3]),
+    type = "l", lty = 1, add = TRUE)
\end{Sinput}
The following code adds the legend at the bottom:
\begin{Sinput}
R> par(mar = rep(0, 4))
R> plot(1, type = 'n', xaxt = 'n', yaxt = 'n',
+    xlab = NA, ylab = NA, frame.plot = FALSE)
R> legend("center", c("I0", "I1", "I2", "I3"),
+    col = 1:4, lwd = 4, horiz = TRUE, bty = "n")
\end{Sinput}

For a numerical comparison, we can calculate the root-mean-squared deviations (RMSDs) between the noisy observations and the reconstructed trajectories at those time points. We can obtain these as follows:

\begin{Sinput}
R> FN.rmsd <- sapply(FNtr, function(x)
+    sqrt(colMeans((subset(x, time %in% FNdat$time) - FNdat[, 2:3])^2))) 
R> colnames(FN.rmsd) <- c("I0", "I1", "I2", "I3")
R> round(FN.rmsd, 3)
\end{Sinput}
\begin{Soutput}
     I0    I1    I2    I3
V 0.234 0.212 0.176 0.167
R 0.281 0.277 0.260 0.255
\end{Soutput}

These results indicate that the estimated parameters fit the observed data better (lower RMSDs) as the denseness of the discretization set is increased. The improvement going from $\bm{I}_2$ and $\bm{I}_3$ is fairly minimal, which confirms that a stable inference result has been achieved and there is no need to further increase the discretization set to $\bm{I}_4$.

\subsection{Setting of hyper-parameters} \label{subsec:phi}

The GP prior on $\bm{x}(t)$ has two ingredients: the mean function $\bm{\mu}(t)$ and covariance kernel/function $\mathcal{K}$. In applications of GPs, it is customary to set $\bm{\mu}(t) = 0$ in the absence of specific prior information \citep[see, e.g., chapter 2 of][]{williams2006gaussian}. The reason this can work well in practice is that the GP, once \textit{conditioned on observations}, will have a mean function that tends to follow the observations. In \pkg{MAGI}, the GP is additionally conditioned on the ODE manifold constraint, which further aligns the GP mean function with the dynamics specified by the ODEs. Thus, our general recommendation is to assume a zero-mean GP prior for simplicity; all of the examples in the paper take this approach and have good inference results. The user may input a custom prior mean function to \pkg{MAGI} by evaluating $\bm{\mu}(\bm{I})$ and $\bm{\dot \mu}(\bm{I})$, i.e., $\bm{\mu}(t)$ and its derivative evaluated at the discretization set $\bm{I}$, and providing them to \code{MagiSolver} via the optional arguments \code{mu} and \code{dotmu}. One potential scenario where this might be helpful is to provide prior guidance for unobserved components, though as demonstrated by the recovery of the unobserved $H$ component in the Hes1 example (Section~\ref{sec:basicusage}), this kind of input is not generally needed.

For a given covariance function $\mathcal{K}$, the GP hyper-parameters $\bm{\phi}$ control the overall prior variance level and prior smoothness/bumpiness of each component's trajectory. Specifically, the default choice and our recommendation for general usage in \pkg{MAGI} is a Matern covariance function $\mathcal{K}$ 
of the form 
\begin{equation}\label{eq:matern}
\mathcal{K}(s, t) = \phi_1\frac{2^{1-\nu}}{\Gamma(\nu)}\left(\sqrt{2\nu}\frac{r}{\phi_2}\right)^\nu B_\nu\left(\sqrt{2\nu}\frac{r}{\phi_2}\right),
\end{equation}
where $r = |s-t|$ is the absolute difference between two time points $s$ and $t$, $\nu = 2.01$ is the smoothness parameter (which ensures twice-differentiable curves),  $\Gamma$ is the gamma function, and $B_\nu$ is the modified Bessel function of the second kind. Larger values of $\phi_1$ therefore favor curves with higher variance, and larger values of $\phi_2$ favor curves with more time-dependence between nearby time points. Each system component has its own set of $(\phi_1, \phi_2)$ values to ensure that the GP has sufficient flexibility to model its dynamics.

Other covariance functions are available in \pkg{MAGI} and may be selected using the \code{kerneltype} optional argument to \code{MagiSolver}. See Appendix \ref{app:covariance} for their specification and details. Note that for a covariance function to be compatible with \pkg{MAGI}, the corresponding GP prior on $\bm{x}(t)$  must satisfy two conditions: (i) the GP derivative $\dot{\bm{x}}(t)$ exists, as implied by the definition of the ODE structure; (ii) $\dot{\bm{x}}(t)$ is also a GP. Together, this requires the covariance function $\mathcal{K}$ to be twice-differentiable (i.e., $\frac{\partial^2}{\partial s\partial t} \mathcal{K}(s, t)$ exists). The number of times that $\mathcal{K}$ is differentiable (with respect to $r$) relates to the smoothness of the GP; higher-order differentiability corresponds to smoother curves. Taking the Matern covariance in Equation~\ref{eq:matern} specifically, $\mathcal{K}$ is $k$-times differentiable if and only if $\nu > k$; thus, smaller values of $\nu$ are more capable of modeling rough or bumpy trajectories \citep[see, e.g., pp.~84-85 of][]{williams2006gaussian}. This motivates our default choice of \code{generalMatern} (i.e.,  Equation~\ref{eq:matern} with $\nu = 2.01$, so that the kernel meets the twice-differentiable requirement and is capable to model relatively rough curves), which in our experience gives the best inference results over the widest range of system behavior (whether rough or smooth). Other covariance functions commonly used in GP applications (such as the Matern covariance with $\nu = 5/2$ or the radial basis function kernel) encode stronger smoothness assumptions that hinder the GP from capturing sharp changes in $\bm{x}(t)$, which may in turn lead to bias in the parameter estimates.\footnote{In general, given that the underlying ODE system can be rough or smooth, it is our experience that using the Matern covariance function with a small value of $\nu$, such as $\nu = 2.01$, provides the GP approximation the capability to model a wide range of dynamic systems (rough or smooth). Our experience suggests that another potential source of bias in \pkg{MAGI} is the fact that the ODE manifold constraint in practical computation can only be applied at a finite set of time points (i.e., $W_{\bm{I}}=0$) rather than on the entire interval (i.e., $W=0$). For systems where $\mathbf{f}$ has very sharp changes, \pkg{MAGI} may still be applicable by choosing a smaller value of the hyper-parameter $\phi_2$ so that those sharp changes can be better approximated by the GP. In this case, visual assessments of the GP fit to the data can provide a good indication of whether subsequent inference with \pkg{MAGI} will be successful. The example in this section demonstrates how to apply this strategy.\label{footnote:smoothing}}

By default, \code{MagiSolver} automatically estimates $\bm{\phi}$ for each system component, depending on the availability of observed data.  Briefly, for components with observations, a GP with the selected covariance function is fitted to the data via \textit{maximum a posteriori} (MAP) estimation with a weakly informative Normal prior for $\phi_2$ (and flat priors otherwise), which provides $\bm{\phi}$ and a value of $\sigma$ to initialize MCMC sampling (if $\sigma$ is unknown). Then to handle unobserved components, optimization is applied to the full posterior in Equation \ref{eq:main} as a function of $\bm{\theta}$ and $\bm{\phi}$,  $\bm{x}(\bm{I})$ for unobserved components, with the previously initialized values of $\sigma$, $\bm{\phi}$, and  $\bm{x}(\bm{I})$ for observed components held fixed.

The values of $\bm{\phi}$ are held fixed during MCMC sampling for $\bm{\theta}$, $\bm{\sigma}$, and $\bm{x}(\bm{I})$. This may be contrasted with a full Bayesian approach for handling $\bm{\phi}$, where $\bm{\phi}$ would also be sampled. MCMC sampling for $\bm{\phi}$ is however expensive as each update requires recomputing the covariance matrices associated with $\bm{x}(\bm{I})$ \citep[e.g., see][]{titsias2011gp} and $\dot{\bm{x}}(\bm{I})$ as needed in \pkg{MAGI}. Thus, we follow the approach of estimating $\bm{\phi}$ based on its marginal likelihood and holding it fixed \citep[e.g., see chapter 5 of][]{williams2006gaussian}; one potential disadvantage is that this approach does not account for uncertainty in $\bm{\phi}$. In practice, credible intervals for $\bm{\theta}$ tend to be fairly stable so long as $\bm{\phi}$ lies within a range that is appropriate for the data; an empirical assessment for this point is provided at the end of this section.

Fixing $\bm{\phi}$ also allows us to leverage techniques to speed up calculations on the covariance matrices associated with the GPs. Specifically, let $\mathcal{K}_d(s, t)$ be the fitted GP covariance function for component $d$, then the following $|\bm{I}| \times |\bm{I}|$ matrices are involved in its GP multivariate normal distribution at the discretization points in $\bm{I}$:
\begin{equation*}
\begin{cases}
C_d &= \mathcal{K}_d(\bm{{I}}, \bm{I}) \\
m_d &= \mathcal{'K}_d(\bm{I}, \bm{I}) \mathcal{K}_d(\bm{I}, \bm{I})^{-1} \\
\Psi_d &= \mathcal{K''}_d(\bm{I}, \bm{I}) - \mathcal{'K}_d(\bm{I}, \bm{I}) \mathcal{K}_d(\bm{I}, \bm{I})^{-1} \mathcal{K'}_d(\bm{I}, \bm{I})
\end{cases}
\end{equation*}
where $\mathcal{'K}_d = \frac{\partial}{\partial s} \mathcal{K}_d(s, t)$, $\mathcal{K'}_d = \frac{\partial}{\partial t} \mathcal{K}_d(s, t)$, and $\mathcal{K''}_d = \frac{\partial^2}{\partial s\partial t} \mathcal{K}_d(s, t)$. Here, $C_d$ is the covariance matrix of $x_d(\bm{I})$, $\Psi_d$ is the covariance matrix of $\dot{x}_d(\bm{I})$, and $m_d$ is the projection matrix that maps $x_d(\bm{I})$ to the mean function of $\dot{x}_d(\bm{I})$. With the GP structure, the covariance between two time points tends to be non-negligible only when they are nearby; thus the off-diagonal entries of $C_d$, $\Psi_d$, and their inverses quickly decay to zero. Likewise, the relation between $x_d(\bm{I})$ is its derivative $\dot{x}_d(\bm{I})$ is local, so off-diagonal entries of $m_d$ also decay to zero quickly. Since $\bm{\phi}$ is fixed, the matrices $C_d^{-1}$, $\Psi_d^{-1}$, and $m_d$ needed in the multivariate normal densities (Equation~\ref{eq:main}) can be pre-computed. Furthermore, we may approximate each of them with sparse band matrices (i.e., non-zero only within \code{bandSize} diagonals on either side of the main diagonal), which reduces the matrix multiplication complexity in Equation~\ref{eq:main} from $O(|\bm{I}|^2)$ to $O(|\bm{I}|)$. This band matrix approximation works best when $\bm{I}$ is evenly-spaced, as recommended in Section~\ref{subsec:discret}. In our experience, a band size of 20 works for most problems and is the default in \pkg{MAGI}. If the approximation fails (i.e., the quadratic form diverges), a warning message that suggests a larger band size will be automatically shown. A different band size can be chosen by setting \code{bandSize} in the \code{control} list to \code{MagiSolver}.

Often, the automatic estimates of $\bm{\phi}$ will be within a reasonable range that permits accurate inference of $\bm{\theta}$; however, this is not guaranteed for all datasets, in which case we may manually override their values for better control over the inference results. This is done by supplying \code{phi} in the \code{control} list to \code{MagiSolver}. \pkg{MAGI} includes the \code{gpsmoothing} function for fitting a GP to data, along with the \code{gpmean} and \code{gpcov} functions to compute the resulting mean vector and covariance matrix conditioned on the data. This allows the user to examine and assess the estimated values of $\bm{\phi}$ prior to running \code{MagiSolver}.

The example in this section demonstrates how these functions can be used, and how the appropriateness of hyper-parameter choices can be assessed, in the context of a system with equations that explicitly depend on time. We consider the three-component system $X = (T_U,T_I,V)$ for HIV infection described in the simulation study of \citet{liang2010estimation}, where $T_U, T_I$ are the concentrations of uninfected and infected cells, and $V$ is the viral load:
\begin{align*}
\mathbf{f}(X, \bm{\theta}, t)  = \begin{pmatrix}
\lambda - \rho T_U - \eta(t) T_U V \\
\eta(t) T_U V - \delta T_I \\
N \delta T_I - c V
\end{pmatrix}. 
\end{align*}
In this system, $\eta(t) = 9\times 10^{-5}  \times (1 - 0.9 \cos(\pi t / 1000))$ is an oscillating infection rate over time (in days), and the parameters to be estimated are $\bm{\theta} = (\lambda, \rho, \delta, N, c)$. Functions for these ODEs (\code{hivtdmodelODE}) and their gradients (\code{hivtdmodelDx} and \code{hivtdmodelDtheta}) are shown in Appendix \ref{app:HIV}. Then the \code{odeModel} list for input to \code{MagiSolver} is as follows:
\begin{Sinput}
R> hivtdmodel <- list(
+    fOde = hivtdmodelODE,
+    fOdeDx = hivtdmodelDx,
+    fOdeDtheta = hivtdmodelDtheta,
+    thetaLowerBound = rep(0, 5),
+    thetaUpperBound = rep(Inf, 5)
+  )
\end{Sinput}
We define the component names and labels for later use:
\begin{Sinput}
R> compnames <- c("TU", "TI", "V")
R> complabels <- c("Concentration", "Concentration", "Load")	
\end{Sinput}
We also create a list with the simulation inputs (parameter values \code{theta}, initial conditions \code{x0}, noise levels \code{sigma}, and observation \code{times}) that mimic those in the referenced paper:
\begin{Sinput}
R> param.true <- list(
+    theta = c(36, 0.108, 0.5, 1000, 3),
+    x0 = c(600, 30, 1e5), 
+    sigma = c(sqrt(10), sqrt(10), 10), 
+    times = seq(0, 20, 0.2) 
+  )
\end{Sinput}
Next we invoke the numerical solver and simulate noisy observations from this system over the specified time interval ($t=0$ to $20$) with measurements at intervals of 0.2 and noise SD \code{param.true$sigma}, again with a random seed set for reproducibility:

\begin{Sinput}
R> set.seed(12321)
R> modelODE <- function(tvec, state, parameters) {
+    list(as.vector(hivtdmodelODE(parameters, t(state), tvec)))
+  }
 
R> xtrue <- deSolve::ode(y = param.true$x0, times = param.true$times,
+                       func = modelODE, parms = param.true$theta)
R> y <- data.frame(xtrue)
R> for(j in 1:(ncol(y) - 1)){
+    y[, 1+j] <- y[, 1+j] + rnorm(nrow(y), sd = param.true$sigma[j])
+  }
\end{Sinput}

\begin{figure}[htbp]
	\centering
	\subcaptionbox{Initial GP fitting to a sample dataset simulated from the HIV model. The black points are the noisy observations. The blue curves represent \pkg{MAGI}'s automatic fit of the GP mean conditional on the data (without ODE information), and the grey bands represent 95\% intervals based on the corresponding GP covariance. The automatically estimated GP hyper-parameters $\phi_1$ and $\phi_2$ allow the curves to follow the observations reasonably well for components $T_U$ and $T_I$; however, the sharp trend in component $V$ is not captured. By using a custom specification of $\phi_1$ and $\phi_2$ for component $V$, the resulting mean curve (orange) can follow the observations well.
		\label{fig:HIVdata-fit}}%
	[\linewidth]{\includegraphics[width=\linewidth]{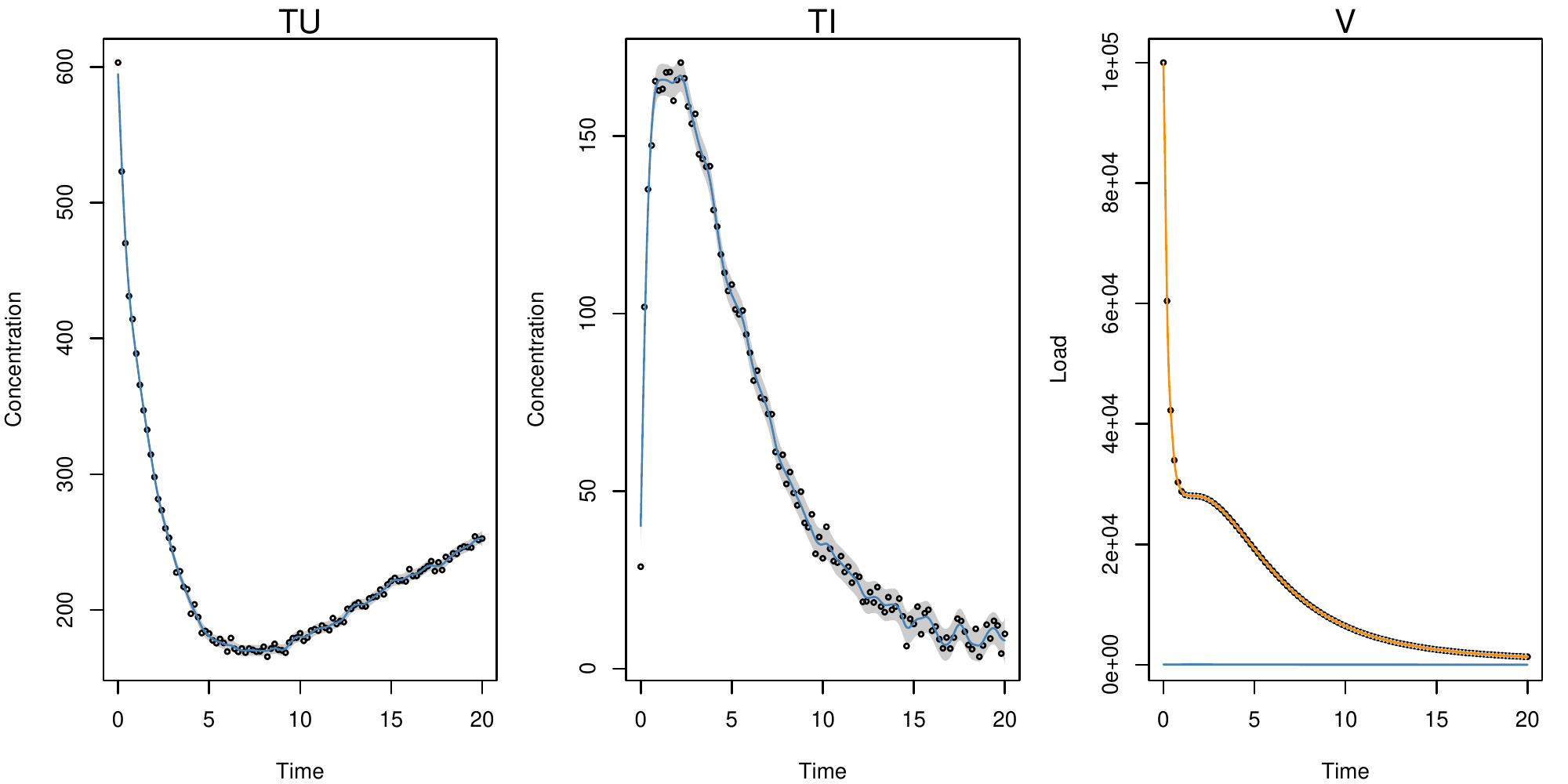}}
	
	\hfill
	\subcaptionbox{Final inferred trajectories (green curves) for the HIV model with 95\% credible bands (light blue). The true trajectories are superimposed in red. Note that the credible bands are very narrow, and are only visible for the $T_I$ component.  \label{fig:HIVinf}}%
	[\linewidth]{\bigskip \includegraphics[width=\linewidth]{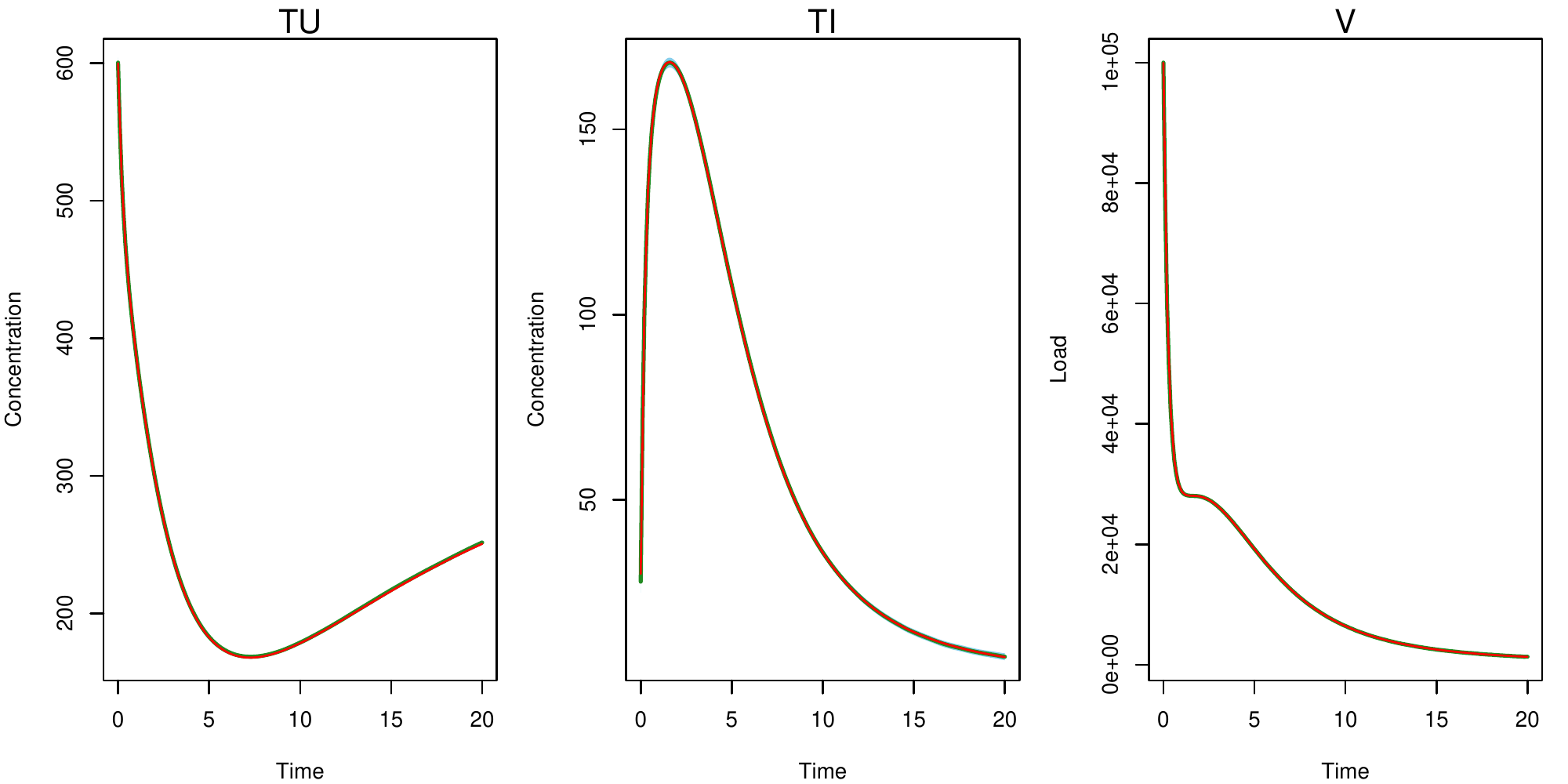}}
	
	\caption{Hyper-parameter setup and inference for a sample dataset simulated from the HIV model. Panel (a) provides a visual assessment of the GP hyper-parameters and their initial fit to the data. Panel (b) shows the final inferred trajectories from \pkg{MAGI}, which closely follow the true trajectories.}
	\label{fig:HIVplots}
\end{figure}

These noisy observations are plotted as the black points in Figure \ref{fig:HIVdata-fit} for each component. The system dynamics are characterized by rapid changes from $t=0$ to $t=5$, with the $V$ component exhibiting a particularly steep decline during the first day.

We can use \code{gpsmoothing} to perform a preliminary GP fit and obtain estimates of $\bm{\phi}$  and $\sigma$ for each component. The inputs to \code{gpsmoothing} are the noisy observations and vector of time points (ODE information is not used at this stage), and the function returns a list with \code{phi} and \code{sigma} elements. Its usage is demonstrated as follows, where we create \code{phiEst} and \code{sigmaInit} to store the results, then perform GP fitting for each system component and extract the estimates:
\begin{Sinput}
R> phiEst <- matrix(0, nrow = 2, ncol = ncol(y) - 1)
R> sigmaInit <- rep(0, ncol(y) - 1)
R> for (j in 1:(ncol(y) - 1)) {
+    hyperparam <- gpsmoothing(y[, j+1], y[, "time"])
+    phiEst[, j] <- hyperparam$phi
+    sigmaInit[j] <- hyperparam$sigma
+  }
\end{Sinput}
Next, we can visualize the GP fit implied by these values of $\bm{\phi}$ and $\sigma$, with the help of the \code{gpmean} and \code{gpcov} functions. These compute the GP mean vector and covariance matrix conditioned on the observations, $\bm{\phi}$, and $\sigma$. To plot a smooth curve of the GP fit, we carry out this computation on a fairly dense set of time points (\code{tOut}), and the diagonal of the covariance matrix can be used to produce credible bands (e.g., $\pm 1.96$ SDs):
\begin{Sinput}
R> tOut <- seq(0, 20, by = 0.025)
R> for (j in 1:3) {
+   plot(y[, "time"], y[, j+1], type = 'n',
+        xlab = "Time", ylab = complabels[j])
+   mtext(compnames[j])
+   fitMean <- gpmean(y[, j+1], y[, "time"], tOut, phiEst[, j], sigmaInit[j])
+   fitCov <- gpcov(y[, j+1], y[, "time"], tOut, phiEst[, j], sigmaInit[j])
+   gp_UB <- fitMean + 1.96 * sqrt(diag(fitCov))
+   gp_LB <- fitMean - 1.96 * sqrt(diag(fitCov))
+   
+   polygon(c(tOut, rev(tOut)), c(gp_UB, rev(gp_LB)),
+           col = "grey80", border = NA)  
+   points(y[, "time"], y[, j+1], cex = 0.5)
+   lines(tOut, fitMean, type = 'l', col = "steelblue")
+ }
\end{Sinput}
The resulting blue curves in Figure \ref{fig:HIVdata-fit} show the GP means of each component, and the grey bands are 95\% intervals  (i.e., $\mbox{mean} \pm 1.96$ SD, where the SDs are extracted from the corresponding GP covariance matrix). We see that the automatically estimated GP hyper-parameters $\phi_1$ and $\phi_2$ allow the mean curves to follow the observations reasonably well for components $T_U$ and $T_I$; however, the sharp trend in component $V$ is not captured. The fitted mean curve for $V$ is close to zero, or equivalently, the observations for $V$ are incorrectly attributed as noise. Recall that at this stage, the mean curves are conditional on the observations only (i.e., ODE manifold constraint is not yet included), so some wiggliness is not unusual, as seen for components $T_U$ and $T_I$. The key visual indicator for a reasonable $\bm{\phi}$ estimate is a mean curve that roughly captures the trajectory implied by the observations. 

The phenomenon seen in the $V$ component might be explained by the fact that GP fitting tends to prefer smoother curves (due to the twice-differentiable requirement, see Section \ref{subsec:phi}), which has the effect here of ``oversmoothing'' the trajectory to a near-constant mean function (see also footnote \ref{footnote:smoothing}). Using these default hyper-parameters could in turn lead to incorrect inference of $\bm{\theta}$. To address this issue, we could (i) choose a smaller value of $\phi_2$ so that the GP can model sharper changes, and (ii) adjust $\phi_1$ according to the overall scale of the component. We examine the automatic estimates of $\bm{\phi}$ and $\bm \sigma$ so that we can make adjustments:
\begin{Sinput}
R> colnames(phiEst) <- compnames
R> phiEst
\end{Sinput}
\begin{Soutput}
              TU           TI            V
[1,] 37538.16828 11083.870501 14299.089897
[2,]     3.91358     2.755717     1.731937
\end{Soutput}
\begin{Sinput}
R> sigmaInit
\end{Sinput}
\begin{Soutput}
[1]     3.378930     3.757803 14158.670863
\end{Soutput}
These estimates appear reasonable for the $T_U$ and $T_I$ components, with a fairly small \code{sigmaInit} that corroborates Figure \ref{fig:HIVdata-fit}, i.e., the GP fit follows the observed data closely. This is further evidenced by the $\phi_1$ values for $T_U$ and $T_I$, where we see that $\sqrt{\phi_1}$ resembles the scale of those components: $\sim$600 for $T_U$ and $\sim$100 for $T_I$. However, the very large value $\sigma\approx 14000 $ for $V$ confirms that the sharp decline in its trajectory is being incorrectly fitted as noise. This is also reflected in its $\bm{\phi}$ estimates:  $\phi_1 \approx 14300$ is too small to adequately capture the variation in $V$, while the bandwidth $\phi_2 \approx 1.7$ is too large to model the sharp initial decline.

Following the strategy described above, we use the \code{control} list to manually specify more suitable values of $\bm{\phi}$ for the $V$ component. In practice, inference is relatively insensitive to $\bm{\phi}$ over a reasonable range of values, so a high level of precision is not required for this step. We increase $\phi_1$ to $10^7$ and decrease $\phi_2$ to 0.5:
\begin{Sinput}
R> phiEst[, 3] <- c(1e7, 0.5)
\end{Sinput}
Since the noise levels $\bm{\sigma}$ are treated as unknown, the values obtained by GP fitting are only used to initialize the MCMC sampler. We may initialize $\sigma$ at a more reasonable (smaller) value for $V$ that corresponds with the update to $\bm{\phi}$, which can also help obtain faster MCMC convergence:
\begin{Sinput}
R> sigmaInit[3] <- 100
\end{Sinput}
To assess whether these manual adjustments to $\bm{\phi}$ and $\sigma$ for $V$ are reasonable, we can recalculate the GP mean and covariance with these new values:
\begin{Sinput}
R> j <- 3
R> fitMean <- gpmean(y[, j+1], y[, "time"], tOut, phiEst[, j], sigmaInit[j])
R> fitCov <- gpcov(y[, j+1], y[, "time"], tOut, phiEst[, j], sigmaInit[j])
\end{Sinput}
We add the updated mean curve (plotted in orange) and 95\% intervals to Figure \ref{fig:HIVdata-fit}:
\begin{Sinput}
R> gp_UB <- fitMean + 1.96 * sqrt(diag(fitCov))
R> gp_LB <- fitMean - 1.96 * sqrt(diag(fitCov))
R> polygon(c(tOut, rev(tOut)), c(gp_UB, rev(gp_LB)),
+          col = "grey80", border = NA)  
R> lines(tOut, fitMean, col = "darkorange")
\end{Sinput}
We see that the mean curve now follows the overall trajectory of the $V$ observations. (The 95\% bands are very narrow and not visible in the plot.) This visually confirms that the new values of $\bm{\phi}$ for component $V$ are reasonable as input to \code{MagiSolver} for carrying out further inference.

We proceed to create a discretization set $\bm{I}$ that adds one equally-spaced time point between observations:
\begin{Sinput}
R> y_I <- setDiscretization(y, level = 1)
\end{Sinput}
Then we run \code{MagiSolver}, supplying \code{phi} and \code{sigma} using the values we specified (recall that \code{phi} is fixed at its supplied value, while \code{sigma} will be sampled via MCMC starting from its supplied value):
\begin{Sinput}
R> HIVresult <- MagiSolver(y_I, hivtdmodel,
+                 control = list(phi = phiEst, sigma = sigmaInit))
\end{Sinput}
We compute posterior means and 95\% credible intervals for $\bm{\theta}$ and $\bm{\sigma}$:
\begin{Sinput}
R> summary(HIVresult, sigma = TRUE, par.names =
+   c("lambda", "rho", "delta", "N", "c", "sigma_TU", "sigma_TI", "sigma_V"))
\end{Sinput}
\begin{Soutput}
      lambda    rho delta   N    c sigma_TU sigma_TI sigma_V
Mean    35.9 0.1070 0.499 958 2.88     3.21     3.41   13.60
2.5%    34.1 0.0982 0.494 943 2.84     2.79     2.95    2.23
97.5%   37.7 0.1150 0.504 973 2.92     3.71     3.96   37.80
\end{Soutput}

The estimates for $\lambda$, $\rho$, $\delta$ are very close to the true values, while $N$ and $c$ have some slight bias. The estimates of $\bm{\sigma}$ likewise reflect their true simulation values, including $\sigma_V$ which we had initialized at 100.  We also extract and plot the inferred trajectories (green) in Figure \ref{fig:HIVinf} together with 95\% credible bands (light blue) and the truth (red) superimposed:
\begin{Sinput}
R> xMean <- apply(HIVresult$xsampled, c(2, 3), mean)
R> xLB <- apply(HIVresult$xsampled, c(2, 3), function(x) quantile(x, 0.025))
R> xUB <- apply(HIVresult$xsampled, c(2, 3), function(x) quantile(x, 0.975))

R> par(mfrow = c(1, 3), mar = c(4, 4, 1.5, 1))
R> for (i in 1:3) {
+    plot(y_I$time, xMean[, i], type = "n", xlab = "Time", ylab = complabels[i])
+    mtext(compnames[i])
+    polygon(c(y_I$time, rev(y_I$time)), c(xUB[, i], rev(xLB[, i])),
+            col = "skyblue", border = NA)  
+    lines(y_I$time, xMean[, i], col = "forestgreen", lwd = 2)
+    lines(param.true$times, xtrue[, i+1], col = "red", lwd = 1)
+  }
\end{Sinput}

In Figure \ref{fig:HIVinf}, the inferred trajectories are indistinguishable from the truth, and the 95\% credible bands are very narrow. The bands are only visible for portions of the $T_I$ trajectory. Thus, good inference results can be obtained in this example by ensuring that the choice of the hyper-parameters $\bm{\phi}$ is reasonable.

\begin{figure}[ht]
	\centering
	\includegraphics[width=\textwidth]{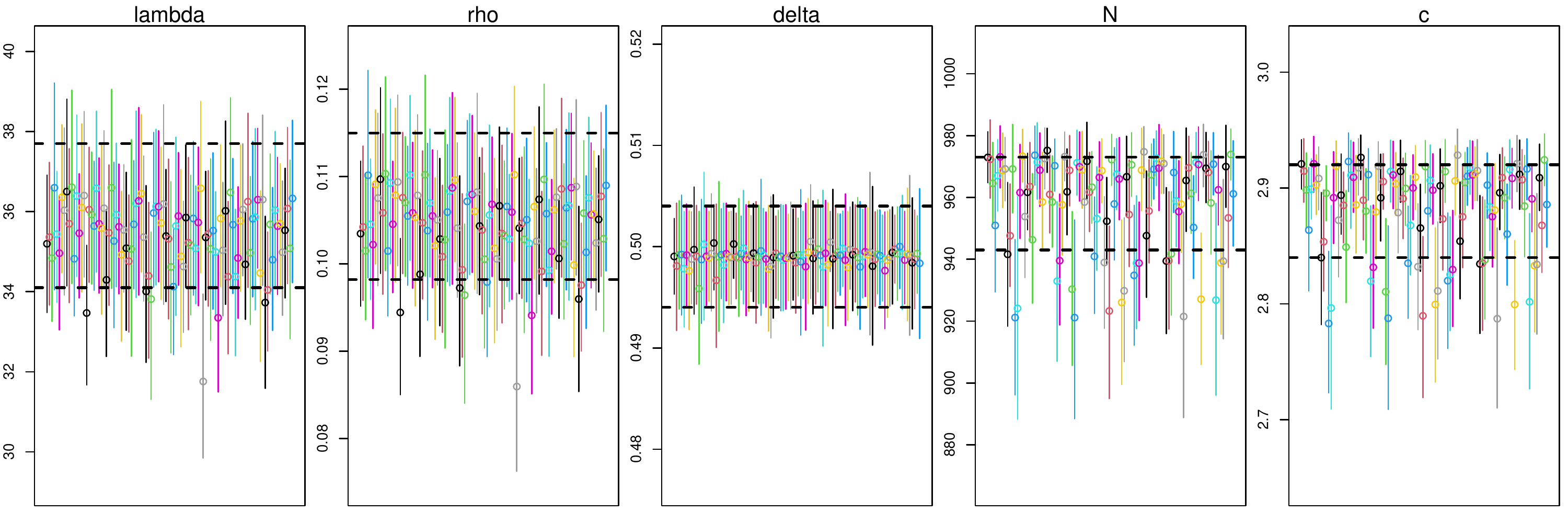}
	\caption{Effect of hyper-parameter values $\bm{\phi}$ on the posterior means (points) and 95\% credible intervals (vertical bars) for each parameter on the HIV simulated dataset. Each bar represents one random set of $\bm{\phi}$ values. The upper and lower limits of the 95\% credible intervals inferred using the original values of $\bm{\phi}$ are shown via the dashed horizontal lines.} \label{fig:HIVsens}
\end{figure}

Lastly, we provide an empirical assessment for the sensitivity of the parameter inference to the GP hyper-parameters $\bm{\phi}$. Since $\bm{\phi}$ is held fixed during MCMC sampling, estimates and credible intervals for $\bm{\theta}$ can have some dependence on the specific value of $\bm{\phi}$ used. In this HIV example, we used $\bm{\phi}$ values that were automatically fit for $T_U$ and $T_I$, and manually specified for the $V$ component. To assess the effect of $\bm{\phi}$, we considered randomly sampling new values for each $\phi_1$ and $\phi_2$, and re-running \code{MagiSolver} with those values. Each entry of $\bm{\phi}$ was drawn uniformly as $[2/3, 3/2]$ times its original value, which could be considered a reasonable range of variation:
\begin{Sinput}
R> for (j in 1:3) {
+    phiEst[, j] <- runif(2, 2/3 * phiEst[, j], 3/2 * phiEst[, j])
+  }
\end{Sinput}
We repeated this for 100 different random seeds. Figure \ref{fig:HIVsens} plots the posterior means (points) and 95\% credible intervals (vertical bars) for the five parameters over all these random $\bm{\phi}$. The upper and lower limits of the 95\% credible intervals inferred using the original values of $\bm{\phi}$ are shown via the dashed horizontal lines. For all five parameters and 99 of 100 repetitions, the 95\% credible intervals overlap with the original ones. The intervals for $\lambda, \rho, \delta$ are especially robust to the choice of $\bm{\phi}$. Overall, this experiment empirically demonstrates fairly stable inference for $\bm{\theta}$, provided that $\bm{\phi}$ lies within a range that is appropriate for the data.

\subsection{Hamiltonian Monte Carlo}\label{subsec:hmc}

HMC is an MCMC sampling algorithm that leverages \textit{Hamiltonian dynamics} \citep[see, e.g.,][]{leimkuhler2004simulating} to obtain draws from a target probability distribution. Via its joint consideration of ``position'' and ``momentum'' variables, samples generated by HMC can explore the target distribution more effectively than those of random walks \citep{neal2011mcmc}. In this subsection, we review the key concepts of HMC and its implementation options available in \pkg{MAGI}.

The ingredients of HMC are setup as follows. Let $\bm{q}$ be a set of ``position'' variables with negative log-density $U(\bm{q})$ (up to an additive constant), so that the target probability density can be written as $\pi(\bm{q}) = \frac{1}{Z}\exp(-U(\bm{q}))$ with $Z$ being the normalizing constant. Under Hamiltonian dynamics, $U(\bm{q})$ is interpreted as the \textit{potential energy} at position $\bm{q}$. Further, let $\nabla U(\bm{q})$ denote the gradient vector of $U(\bm{q})$, with respect to $\bm{q}$. In \pkg{MAGI}, $\bm{q}$ consists of all the variables to be sampled: the trajectories $\bm{x}(\bm{I})$ and the parameters $\bm{\theta}$ (which also includes the noise levels $\bm{\sigma}$ if they are unknown); the function $U(\cdot)$ is the negative log of their joint posterior density, as shown in Equation~\ref{eq:main}. Next, HMC introduces ``momentum'' variables $\bm{p}$ with the same dimension as $\bm{q}$, and we define their \textit{kinetic energy} as $K(p) = \bm{p}^\top \bm{p} /2$. The \textit{Hamiltonian} then combines the kinetic and potential energies, defined as $H(\bm{q}, \bm{p}) = U(\bm{q}) + K(\bm{p})$. 

With this setup, $\exp[-H(\bm{q}, \bm{p})]$ specifies a joint density (up to a multiplicative constant) of $\bm{q}$ and $\bm{p}$, which can be interpreted as follows: (i) $\bm{q}$ and $\bm{p}$ are independent random variables; (ii) the marginal distribution of $\bm{q}$ is the target posterior of interest in Equation~\ref{eq:main}; (iii) the marginal distribution of $\bm{p}$ is multivariate standard normal. Intuitively, HMC works on this joint density so that we obtain the samples of interest for $\bm{q}$, and the role of $\bm{p}$ is to facilitate the sampling efficiency.

From a current state for $\bm{q}$, one iteration of the HMC algorithm generates the next state for $\bm{q}$ as follows, where $L$ is a positive integer and $\bm{\epsilon} > 0$ is a vector of step sizes: 
\begin{enumerate}
	\item Draw $\bm{p}$ from a standard multivariate normal distribution, then set $(\bm{q}_0, \bm{p}_0) = (\bm{q},\bm{p})$.
	\item For $l = 1, \ldots, L$, the \textit{leapfrog method} is used to approximate the Hamiltonian dynamics:
	\begin{enumerate}
		\item Take a half-step for the momentum by setting $\tilde{\bm{p}} = \bm{p}_{l-1} - (\bm{\epsilon} /2) \cdot \nabla U(\bm{q}_{l-1})$.
		\item Take a full-step for the position by setting $\bm{q}_l = \bm{q}_{l-1} + \bm{\epsilon} \cdot \tilde{\bm{p}}$.\footnote{When $\bm{q}$ has upper or lower bounds (e.g., parameters that are strictly positive), this step is modified slightly to handle the bounds. For details see p.~149 of \cite{neal2011mcmc}.}
		\item Take a half-step for the momentum (using the updated position) by setting $\bm{p}_l = \bm{p}_{l-1} - (\bm{\epsilon}/2)  \cdot  \nabla U(\bm{q}_{l})$.
	\end{enumerate}
	\item The proposed state is  $(\bm{q}^*, \bm{p}^*) \equiv (\bm{q}_L, \bm{p}_L)$. Accept $\bm{q}^*$ as the next state of $\bm{q}$ with the usual Metropolis acceptance probability, i.e., $\min[1, \exp(-H(\bm{q}^*, \bm{p}^*) + H(\bm{q}, \bm{p}))]$. If the proposed state is rejected, the next state of $\bm{q}$ is set to be the same as its current state.
\end{enumerate}
Two main aspects of the HMC algorithm can thus be tuned: the step size vector $\bm{\epsilon}$, and the number of leapfrog steps $L$. In \pkg{MAGI}, these can be supplied by the user in the list of optional inputs to \code{MagiSolver}: $L$ is specified via \code{nstepsHmc} and a starting factor for $\bm{\epsilon}$ is specified via \code{stepSizeFactor}. Two other options in \code{MagiSolver} related to MCMC sampling are also worth mentioning: the number of HMC iterations to run is specified via \code{niterHmc}, and the proportion of MCMC samples to discard as an initial burn-in period is specified via \code{burninRatio}. We discuss each of these and our practical recommendations below.

The step-size vector $\bm{\epsilon}$ controls the accuracy of the Hamiltonian approximation: if $\bm{\epsilon}$ is too large, the HMC proposals will have a high rejection rate; while if $\bm{\epsilon}$ is too small, the HMC proposals will move slowly around the target posterior distribution.
Thus $\bm{\epsilon}$ needs to be carefully tuned to achieve good HMC performance. As suggested in \cite{neal2011mcmc}, the optimal acceptance rate of HMC is 65\%; moreover, the tuning of $\bm{\epsilon}$ can be done independently of $L$. \pkg{MAGI} handles these aspects of $\bm{\epsilon}$ via automatic tuning during the burn-in period (i.e., the first \code{burninRatio * niterHmc} iterations). Briefly, \pkg{MAGI} uses a moving window of 100 iterations to monitor: (i) the acceptance rate, so that $\bm{\epsilon}$ is increased (or decreased) if the acceptance rate is above 90\% (or below 60\%); (ii) the SD of each variable in $\bm{q}$, so that the individual step sizes in $\bm{\epsilon}$ are adapted to follow the scale of each variable.\footnote{\cite{neal2011mcmc} recommends randomizing the step size for each HMC iteration to further improve the stability of HMC; so in practice, at each iteration \pkg{MAGI} draws a random step size vector uniformly from the range $[\bm{\epsilon}, 2\bm{\epsilon}]$.} The optional \code{stepSizeFactor} input can be a scalar (applied to all the variables in $\bm{q}$) or a vector (with the same length as $\bm{q}$), that specifies the starting value of $\bm{\epsilon}$. In our experience, \pkg{MAGI}'s automatic tuning can quickly identify the range of $\bm{\epsilon}$ needed for efficient HMC sampling, so that there is usually no need to manually specify \code{stepSizeFactor}. If the acceptance rate of the first many (e.g., thousands) HMC iterations is 0\% or 100\%, overriding the default \code{stepSizeFactor = 0.01} with a value that is several orders of magnitude smaller (if the acceptance rate is 0\%) or larger (if the acceptance rate is 100\%) could help speed up convergence.

Since automatic tuning of $\bm{\epsilon}$ is limited to the burn-in period, the default \code{burninRatio = 0.5} usually provides a good balance between samples used for convergence/tuning and samples for final inference. The number of HMC iterations (default \code{niterHmc = 20000}) can be set to balance computational time constraints with the Monte Carlo variance of the resulting estimates. To monitor sampling progress when running \code{MagiSolver}, setting \code{verbose = TRUE} will print a message to the console every 100 HMC iterations.

\begin{figure}[ht]
	\centering
	\includegraphics[width=5.5in]{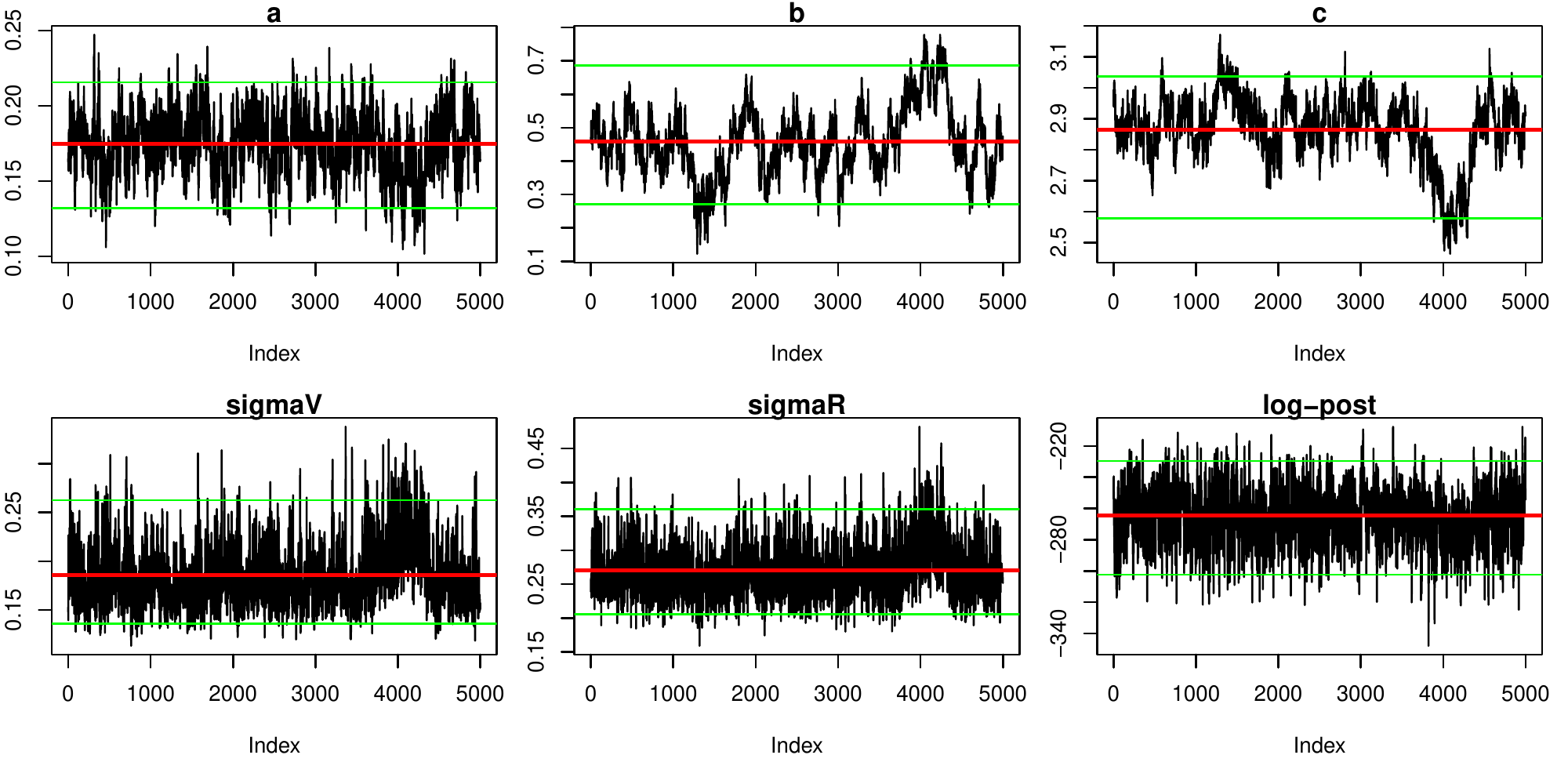}
	\caption{MCMC traceplots that indicate relatively high autocorrelation, based on an example run for the FN system with a dense discretization set and 200 leapfrog steps per HMC iteration. The horizontal lines in the plots indicate the posterior mean (red) and limits of the 95\% credible interval (green) for each parameter.} \label{fig:fnstickytrace}
\end{figure}

In \pkg{MAGI}, the number of leapfrog steps $L$ is fixed for all HMC iterations. The default value of \code{nstepsHmc} is 200 and should work well in many cases.\footnote{Since the dimensionality of $\bm{q}$ in \pkg{MAGI} includes all the discretization points $\bm{x}(\bm{I})$, which is often fairly large, we use a more conservative default of $L=200$ compared to the $L=100$ rule-of-thumb suggested by \cite{neal2011mcmc}.} We recommend using traceplots (which can be generated via the convenience \code{plot()} function on the output of \code{MagiSolver}) to visually diagnose whether a larger $L$ might be needed. The overall cost per HMC iteration is roughly proportional to $L$, so $L$ should only be increased if necessary. As the discretization set $\bm{I}$ is taken to be increasingly dense (see Section \ref{subsec:discret}), the increased dimensionality of $\bm{q}$ may lead to `sticky' HMC samples with high autocorrelation. To illustrate, let us revisit the FN dataset and run \code{MagiSolver} on \code{y_I3} (which has 321 discretization points per component) with the default $L=200$:
\begin{Sinput}
R> data("FNdat")
R> set.seed(12321)
R> y_I0 <- setDiscretization(FNdat, by = 0.5)
R> y_I3 <- setDiscretization(y_I0, level = 3)
R> FNres3b <- MagiSolver(y_I3, fnmodel, control = list(niterHmc = 10000))	
\end{Sinput}
Then, we examine the traceplots of the parameters, shown in Figure \ref{fig:fnstickytrace}:
\begin{Sinput}
R> plot(FNres3b, type = "trace", par.names = FNpar.names, sigma = TRUE)
\end{Sinput}

We see that the MCMC samples exhibit some non-stationary patterns, rather than appearing as random scatters for each parameter. This is most evident for parameters $b$ and $c$, where the Markov chain can remain in one region of their distribution for several hundred iterations at a time (hence `sticky'). This issue can usually be alleviated by increasing $L$, for example by running \code{MagiSolver} with \code{nstepsHmc = 1000} as in Section \ref{subsec:discret}, then the re-drawn traceplots can be seen to indicate good convergence (not shown for brevity).

\section{Benchmark comparisons with other methods}\label{sec:comparisons}

As noted in the Introduction, unobserved system components pose a challenge to most other methods of ODE inference and their software implementations. In this section, we take the Hes1 example where component $H$ is unobserved (as presented in the Introduction and Section \ref{sec:basicusage}), and compare the inference accuracy and runtime of different software packages that can handle this problem in \proglang{R}.

Alongside \pkg{MAGI}, we consider the \pkg{deBInfer} and \pkg{CollocInfer} packages (see Section \ref{sec:related}). The \pkg{deBInfer} package represents a Bayesian approach to parameter estimation with the help of numerical ODE solvers, and hence is generally applicable to systems with unobserved components. The \pkg{CollocInfer} package is a collocation-based penalized likelihood method that uses a B-spline basis. While \pkg{CollocInfer} (along with \pkg{pCODE}, which is based on the same underlying methodology) can be used to perform inference with an unobserved component, an estimate of the B-spline basis must be supplied by the user, as we detail below. The other \proglang{R} packages described in Section \ref{sec:related} do not have the capacity for unobserved components.

To generate simulated datasets for this comparison, we follow the same procedure described in Section \ref{sec:hes1intro}, with 100 different random seeds. Since \pkg{deBInfer} and \pkg{CollocInfer} do not directly provide estimates of the inferred trajectories, for fair comparison we use each method to obtain estimates of the parameters $\bm{\theta} = (a,b,c,d,e,f,g)$ and initial conditions $P(0), M(0), H(0)$. Using these estimated parameters and initial conditions, we run the numerical solver to reconstruct the trajectory implied by the estimates. For a given method, we then compute the RMSE between this reconstructed trajectory and the true trajectory (i.e., the solid curves in Figure \ref{fig:hes1obs}) for each system component, at the 33 observation time points. We call this the \textit{trajectory RMSE} metric, as in \citet{yang2021inference}.

Thus, on each simulated dataset, the three methods are run as follows to obtain estimates of $\bm{\theta}$ and $P(0), M(0), H(0)$:
\begin{itemize}
\item \pkg{MAGI} is run as described in Section \ref{sec:basicusage}, which infers the parameters and system trajectories. The posterior means of $\bm{\theta}$ and $P(0), M(0), H(0)$ from the \code{MagiSolver} output are taken as the estimates.
\item For \pkg{deBInfer}, we set up a normal likelihood for the observations of $P$ and $M$ on the log-scale, with known SD from the simulation setup. The priors for $\bm{\theta}$ were set to be uniform over restricted ranges: $a,b,c,d,e$ are taken to be uniform on $[0,2]$, $f$ is uniform on $[0,100]$, and $g$ is uniform on $[0,10]$. Without imposing these informative priors on $\bm{\theta}$ (i.e., which contrast with the uniform priors over all positive real numbers, as used in \pkg{MAGI}), the method would often fail to converge at a reasonable result. The priors for the initial conditions were uniform on the log-scale. We ran 20000 MCMC iterations, to match the number of iterations run in \pkg{MAGI}. The posterior means of $\bm{\theta}$ and $P(0), M(0), H(0)$ from the \code{de_mcmc} output are taken as the estimates, after discarding the first 10000 iterations as burn-in.
\item For \pkg{CollocInfer}, the method begins by computing initial estimates of the B-spline basis. However, the package does not have the ability to compute such estimates when there are unobserved system components. Therefore, manual input is needed in this case to supply these estimates, which we do as follows. First, we use the package functions to fit the B-spline basis given the \textit{true} values of all the system components at the 33 observation time points. (In real data analyses, such true values would not be available, so \pkg{CollocInfer} is given an additional advantage by taking this approach.) Second, we take the B-spline fit for the unobserved component obtained from the first step, together with the actual noisy observations of $P$ and $M$, and run the main \pkg{CollocInfer} method with its default settings to obtain the final estimates of $\bm{\theta}$ and $P(0), M(0), H(0)$.
\end{itemize}
The full code to benchmark each method is provided in the replication materials.

The results of the three methods over the 100 simulated datasets are summarized in Table \ref{tab:comparison}. For each method, the average runtime and trajectory RMSE of each system component is shown. For the unobserved $H$ component, \pkg{MAGI} can reliably recover its trajectory, while \pkg{deBInfer} and \pkg{CollocInfer} cannot, as indicated by the large RMSEs. For the protein ($P$) and mRNA ($M$) components, where observations are available, the results are relatively closer, though \pkg{MAGI} still outperforms the other two methods. In terms of runtime, \pkg{MAGI} is the fastest of the three methods, averaging 6.2 minutes per dataset. \pkg{CollocInfer}, as a frequentist-based optimization method, is also relatively fast; \pkg{deBInfer}, which relies on the numerical solver at each iteration to compute the likelihood, was significantly slower than the other two methods. All of the computations were carried out using a single CPU core of an Intel Xeon 3.7 GHz processor. Overall, these results illustrate the favorable performance of \pkg{MAGI} on this inference problem.

\begin{table}[ht]
	\centering
	\begin{tabular}{cccc}
		\hline
		 & \pkg{MAGI} & \pkg{deBInfer}  & \pkg{CollocInfer} \\ 
		\hline
Trajectory RMSE of $P$ & 0.95 & 1.36 & 1.49 \\
Trajectory RMSE of $M$ & 0.20 & 0.33 & 0.45 \\
Trajectory RMSE of $H$ & 2.48 & 8.39 & 226.0 \\
Runtime (minutes) & 6.2 & 46.7 & 7.4 \\
		\hline
	\end{tabular}
	\caption{Performance of the three compared methods over 100 simulated datasets from the Hes1 model. For each method, the mean runtime and trajectory RMSE of each component is reported. The $H$ component is never observed.}\label{tab:comparison}
\end{table}

\section{Conclusion and discussion}

The inference problem for dynamic systems is a vital task in science and engineering, which widely use ODE models. In practice, the experimental data collected from these systems may often be noisy and sparse. Furthermore, some components of the system may be entirely unobserved. These features pose challenges for estimating the unknown parameters and reconstructing the system trajectories without numerical integration. Existing software packages for this task, to the best of our knowledge, cannot readily handle unobserved components without substantial manual input. This paper presented our package for the \pkg{MAGI} method \citep{yang2021inference}, which capably handles these inference problems in a principled Bayesian framework using manifold-constrained Gaussian processes. The user may choose any of \proglang{R}, \proglang{MATLAB}, and \proglang{Python} to carry out the analyses. Scripts that demonstrate the equivalent functionality in all three environments are included in the replication materials.

We believe the methodological approach of \pkg{MAGI} is quite extensible. We discuss some interesting directions for future developments to the software package in the following.
\begin{itemize}
\item Extending the inference framework to allow for time-varying parameters. The current implementation of \pkg{MAGI} assumes time-constant parameters $\bm{\theta}$, which covers a broad range of dynamic systems. In some cases, a more flexible time-varying specification $\bm{\theta}(t)$ is needed, e.g., for pharmacokinetic parameters \citep{li2002estimation} and disease transmission rates \citep{keller2022tracking}. GP priors for $\bm{\theta}(t)$ could potentially be incorporated into the \pkg{MAGI} framework as well to handle this situation.
\item Incorporating more flexible choices for the measurement error model and priors for the parameters. The \pkg{MAGI} package currently assumes the noise term can be adequately modeled as  additive and Gaussian. Multiplicative noise can be handled by taking a log-transformation, as demonstrated in the Hes1 example. If the measurement noise is significantly non-Gaussian, allowing a custom specification for the likelihood $ p(  \bm{y}(\bm \tau) | \bm{x}(\bm{I}))$ could be a useful feature. Further, the priors for $\bm{\theta}$ are assumed to be flat (uniform) between the user-provided bounds \code{thetaLowerBound} and \code{thetaUpperBound}. While this assumption may often be adequate (as transformations may also be applied to parameters as needed), custom priors $\pi(\bm{\theta})$ could allow more specific prior knowledge of the parameters to be incorporated into the inference.
\item Functions to help set up the ODE gradients with respect to $\bm{\theta}$ and $X$. Supplying the analytic gradients $\partial \mathbf{f}(X, \bm{\theta}, t)/\partial X$ and $\partial \mathbf{f}(X, \bm{\theta}, t)/\partial \bm{\theta}$ enables \pkg{MAGI} to draw MCMC samples efficiently via its  HMC implementation for the target posterior density. The ability to easily generate code for these gradients from the ODEs (e.g., with the help of symbolic differentiation) or have them automatically computed (i.e., automatic differentiation or \textit{autograd}) could potentially reduce the time needed to set up a dynamic system in \pkg{MAGI}. In \proglang{R}, support for autograd is currently experimental and computationally inefficient compared with supplying the analytic gradients; a proof-of-concept that uses autograd with \pkg{MAGI} is provided in Appendix \ref{app:autodiff} for the interested reader.
\end{itemize}

\bibliography{jss4695}

%
%
%
\newpage

\begin{appendix}

\section[MAGI usage in MATLAB]{\pkg{MAGI} usage in \proglang{MATLAB}} \label{app:matlab}

To use \pkg{MAGI} in \proglang{MATLAB}, first clone the repository at \url{https://github.com/wongswk/magi}, e.g., by executing on the command-line \code{git clone https://github.com/wongswk/magi}. On Linux-compatible systems, build the core \proglang{C++} \pkg{MAGI} library by running the \code{build.sh} shell script, then run the \code{MATLAB_build.sh} shell script in the \code{MATLABmagi} directory to generate the compiled  \proglang{MEX} files. For Windows systems, pre-compiled versions of the \pkg{MAGI} library and \proglang{MEX} files are provided in the \code{MATLABmagi/windows} directory. Ensure that \code{libcmagi.so} (or \code{libcmagi.dll} in Windows), along with the accompanying \code{.m} routines and compiled \proglang{MEX} files of the package, are located in either the working directory or the path of \proglang{MATLAB}.

The replication package (supplementary .zip file provided with this article) includes the \proglang{MATLAB} directory which contains the materials for  running the three examples discussed in this paper in \proglang{MATLAB}. Specifically:
\begin{itemize}
	\item The \code{models/} subdirectory contains the functions for the ODE systems, since the \proglang{MATLAB} convention is one function per file. For example, for the FitzHugh-Nagumo (FN) equations,
	\begin{itemize}
		\item \code{fnmodelODE.m} codes the ODEs,
		\item \code{fnmodelDx.m} codes their gradients with respect to $X$
		\item \code{fnmodelDtheta.m} codes their gradients with respect to $\bm{\theta}$
 		\item \code{fnmodelODEsolve.m} codes the system in a form suitable for invoking ODE solvers such as \code{ode45}.
	\end{itemize}
	\item \code{replication.m} is the \proglang{MATLAB} replication script that carries out the same analyses described in the main paper. Please note that the random seeds between \proglang{R} and \proglang{MATLAB} are not interchangeable so their corresponding numerical results are expected to have slight differences attributable to random-number generation. Otherwise, the functionalities of the \proglang{R} and \proglang{MATLAB} replication scripts are identical, so that users can easily follow the equivalent syntax in \proglang{MATLAB} for the examples given in this paper.
\end{itemize}

We briefly point out two main differences of note between the syntax in \proglang{R} and \proglang{MATLAB}:
\begin{itemize}
	\item In \proglang{MATLAB}, \code{struct} arrays are used in place of \code{list} objects in \proglang{R}. Taking the FN equations as an example, to set up \code{odeModel} for input to \code{MagiSolver} we create a \code{struct} with the five required elements, where \code{fOde}, \code{fOdeDx}, and \code{fOdeDtheta} are assigned their corresponding function handles:
\begin{Sinput}
>> fnmodel.fOde = @fnmodelODE;
>> fnmodel.fOdeDx = @fnmodelDx;
>> fnmodel.fOdeDtheta = @fnmodelDtheta;
>> fnmodel.thetaLowerBound = [0 0 0];
>> fnmodel.thetaUpperBound = [Inf Inf Inf];	
\end{Sinput}
Similarly, the \code{control} list is set up by creating a \code{struct} in \proglang{MATLAB}, e.g., 
\begin{Sinput}
>> config.niterHmc = 10000;
\end{Sinput}
to run 10000 HMC iterations, and then \code{config} can be passed as the \code{control} argument to \code{MagiSolver}.

The output of \code{MagiSolver} is also a \code{struct}; e.g., if \code{FNres0} contains the output of a \code{MagiSolver} run, the matrix of MCMC samples for $\bm{\theta}$ would be accessed as \code{FNres0.theta}. The convenience functions \code{summaryMagiOutput()} and \code{plotMagiOutput()} are provided (analogous to \code{summary()} and \code{plot()} in \proglang{R}) to generate a table of parameter estimates (with credible intervals) and visualize the inferred trajectories from the output of \code{MagiSolver}.

\item In \proglang{MATLAB}, optional function arguments need to be explicitly skipped by passing \code{[]}. For example, \code{MagiSolver} allows the time vector to be passed as a separate argument from the data matrix \code{y}.  When the time vector is included as the first column in \code{y}, we would skip the third argument as follows:
\begin{Sinput}
>> FNres0 = MagiSolver(y, fnmodel, [], config);
\end{Sinput}

\end{itemize}

\section[MAGI usage in Python]{\pkg{MAGI} usage in \proglang{Python}} \label{app:python}

To use \pkg{MAGI} in \proglang{Python}, first clone the repository at \url{https://github.com/wongswk/magi}, e.g., by executing on the command-line \code{git clone https://github.com/wongswk/magi}. Build the core \proglang{C++} \pkg{MAGI} library by running the \code{build.sh} shell script, then run the \code{py_build.sh} shell script in the \code{pymagi} directory to build the \code{pymagi.so} Python library. Ensure that this \code{pymagi} directory is contained in \proglang{Python}'s path.

The replication package (supplementary .zip file provided with this article) includes the \code{python} directory which contains the \proglang{Python} script \code{replication.py} that carries out the same analyses for the three examples discussed in the main paper. Please note that the random seeds between \proglang{R} and \proglang{Python} are not interchangeable so their corresponding numerical results are expected to have slight differences attributable to random-number generation. Otherwise, the functionalities of the \proglang{R} and \proglang{Python} replication scripts are identical, so that users can easily follow the equivalent syntax in \proglang{Python}.

We briefly point out two main differences of note between the syntax in \proglang{R} and \proglang{Python}:

\begin{itemize}
	\item In \proglang{Python}, we construct the \code{odeModel} input to \code{MagiSolver} by using the helper function \code{ode_system}, rather than setting up an \proglang{R} \code{list}.  Taking the FN equations as an example, suppose \code{fnmodelOde}, \code{fnmodelDx}, and \code{fnmodelDtheta} respectively are the functions for the ODEs, gradients with respect to $X$, and gradients with respect to $\bm{\theta}$. Then we can call \code{ode_system} as follows:
\begin{Sinput}
>> fn_system = ode_system("FN-python",
                  fnmodelOde, fnmodelDx, fnmodelDtheta,
                  thetaLowerBound = np.array([0, 0, 0]),
                  thetaUpperBound = np.array([np.inf, np.inf, np.inf]))
\end{Sinput}
where the first argument can be any string that provides a name for the system.

	\item In \proglang{Python}, the dictionary data type (\code{dict}) is used in place of the \proglang{R} \code{list} for the \code{control} argument to \code{MagiSolver}. For example, with \code{fn_system} as defined above, we can provide the \code{control} argument in the call to \code{MagiSolver} as follows:
	\begin{Sinput}
>> FNres3 = MagiSolver(y = y_I3, odeModel = fn_system,
                control = dict(niterHmc = 10000, nstepsHmc = 1000))
	\end{Sinput}
The output of \code{MagiSolver} is also a \code{dict}; e.g., if \code{FNres0} contains the output of a \code{MagiSolver} run, the matrix of MCMC samples for $\bm{\theta}$ would be accessed as	\begin{Sinput}
>> FNres0['theta']
\end{Sinput}
The convenience functions \code{summaryMagiOutput()} and \code{plotMagiOutput()} are provided (analogous to \code{summary()} and \code{plot()} in \proglang{R}) to generate a table of parameter estimates (with credible intervals) and visualize the inferred trajectories from the output of \code{MagiSolver}.
\end{itemize}

\section{Functions for Fitzhugh-Nagumo ODEs and their gradients} \label{app:FN}

The following \proglang{R} functions encode the ODEs and gradients for the FN equations discussed in Section \ref{subsec:discret}.

\begin{Sinput}
R> fnmodelODE <- function(theta, x, tvec) {
+    V <- x[, 1]
+    R <- x[, 2]
+   
+    result <- array(0, c(nrow(x), ncol(x)))
+    result[, 1] = theta[3] * (V - V^3 / 3.0 + R)
+    result[, 2] = -1.0/theta[3] * (V - theta[1] + theta[2] * R)
+   
+    result
+  }

R> fnmodelDx <- function(theta, x, tvec) {
+    resultDx <- array(0, c(nrow(x), ncol(x), ncol(x)))
+    V = x[, 1]
+   
+    resultDx[, 1, 1] = theta[3] * (1 - V^2)
+    resultDx[, 2, 1] = theta[3]
+   
+    resultDx[, 1, 2] = -1.0 / theta[3]
+    resultDx[, 2, 2] = -theta[2] / theta[3]
+   
+    resultDx
+  }
	
R> fnmodelDtheta <- function(theta, x, tvec) {
+    resultDtheta <- array(0, c(nrow(x), length(theta), ncol(x)))
+   
+    V = x[, 1]
+    R = x[, 2]
+   
+    resultDtheta[, 3, 1] = V - V^3 / 3.0 + R
+   
+    resultDtheta[, 1, 2] =  1.0 / theta[3] 
+    resultDtheta[, 2, 2] = -R / theta[3]
+    resultDtheta[, 3, 2] = 1.0 / (theta[3]^2) * (V - theta[1] + theta[2] * R)
+   
+    resultDtheta
+  }
\end{Sinput}

\section{Functions for HIV model ODEs and their gradients} \label{app:HIV}

The following \proglang{R} functions encode the ODEs and gradients for the HIV model discussed in Section \ref{subsec:phi}.

\begin{Sinput}
R> hivtdmodelODE <- function(theta, x, tvec) {
+    TU <- x[, 1]
+    TI <- x[, 2]
+    V <- x[, 3]
+   
+    lambda <- theta[1]
+    rho <- theta[2]
+    delta <- theta[3]
+    N <- theta[4]
+    c <- theta[5]
+   
+    eta <- 9e-5 * (1 - 0.9 * cos(pi * tvec / 1000))
+   
+    result <- array(0, c(nrow(x), ncol(x)))
+    result[, 1] = lambda - rho * TU - eta * TU * V
+    result[, 2] = eta * TU * V - delta * TI
+    result[, 3] = N * delta * TI - c * V
+   
+    result
+  }

R> hivtdmodelDx <- function(theta, x, tvec) {
+    resultDx <- array(0, c(nrow(x), ncol(x), ncol(x)))
+   
+    TU <- x[, 1]
+    TI <- x[, 2]
+    V <- x[, 3]
+   
+    lambda <- theta[1]
+    rho <- theta[2]
+    delta <- theta[3]
+    N <- theta[4]
+    c <- theta[5]
+   
+    eta <- 9e-5 * (1 - 0.9 * cos(pi * tvec / 1000))
+   
+    resultDx[, , 1] = cbind(-rho - eta * V, 0, -eta * TU)
+    resultDx[, , 2] = cbind(eta * V, -delta, eta * TU)
+    resultDx[, , 3] = cbind(rep(0, nrow(x)), N * delta, -c)
+ 
+    resultDx
+  }

R> hivtdmodelDtheta <- function(theta, x, tvec) {
+    resultDtheta <- array(0, c(nrow(x), length(theta), ncol(x)))
+   
+    TU <- x[, 1]
+    TI <- x[, 2]
+    V <- x[, 3]
+   
+    delta <- theta[3]
+    N <- theta[4]
+ 
+    resultDtheta[, , 1] = cbind(1, -TU, 0, 0, 0)
+    resultDtheta[, , 2] = cbind(0, 0, -TI, 0, 0)
+    resultDtheta[, , 3] = cbind(0, 0, N * TI, delta * TI, -V)
+ 
+    resultDtheta
+  }
\end{Sinput}

\section[Combining automatic differentiation with MAGI]{Combining automatic differentiation with \pkg{MAGI}}\label{app:autodiff}

We can utilize automatic differentiation (autograd) with \pkg{MAGI} by leveraging the autograd functionality provided by the \pkg{torch} package \citep{torch_Rpack}. First, we install the \pkg{torch} package in \proglang{R} by executing the command \code{install.packages("torch")}.

To integrate autograd with \pkg{MAGI}, it is necessary to rewrite the ODE function using torch tensors instead of \proglang{R} arrays. For the Hes1 example presented in Section \ref{sec:basicusage}, a single line of code needs to be modified: the \proglang{R} array \code{PMHdt = array(0, c(nrow(x), ncol(x)))} is replaced with the torch tensor \code{PMHdt = torch_empty(dim(x))}. An adapted version of \code{hes1logmodelODE} suitable for autograd is as follows:

\begin{Sinput}
R> hes1logmodelODE_torch <- function (theta, x, tvec) {
+   P = exp(x[, 1])
+   M = exp(x[, 2])
+   H = exp(x[, 3])
+   
+   PMHdt <- torch_empty(dim(x))
+   PMHdt[, 1] = -theta[1] * H + theta[2] * M / P - theta[3]
+   PMHdt[, 2] = -theta[4] + theta[5] / (1 + P^2) / M
+   PMHdt[, 3] = -theta[1] * P + theta[6] / (1 + P^2) / H - theta[7]
+   
+   PMHdt
+ }
\end{Sinput}

With this new implementation of the \code{hes1logmodelODE_torch} function, the following function can be used to calculate the derivatives with respect to both $X$ and $\bm{\theta}$:

\begin{Sinput}
R> ode_autograd <- function(ode_func_torch, theta, x, tvec) {
+   # Convert input arguments to torch tensors with requires_grad = TRUE
+   theta <- torch_tensor(theta, requires_grad = TRUE)
+   x <- torch_tensor(x, requires_grad = TRUE)
+   tvec <- torch_tensor(tvec)
+ 
+   # Calculate output using torch operations
+   output = ode_func_torch(theta, x)
+   
+   # Initialize gradient matrices
+   ode_dtheta = array(dim=c(nrow(output), length(theta), ncol(output)))
+   ode_dx = array(dim=c(nrow(output), ncol(output), ncol(output)))
+   
+   # Calculate gradients for each element in the output
+   for (i in 1:nrow(output)) {
+     for (j in 1:ncol(output)) {
+       # Zero out gradients from previous iterations
+       if (length(theta$grad) > 0) {
+         theta$grad$zero_()
+       }
+       if (length(x$grad) > 0) {
+         x$grad$zero_()
+       }
+       
+       # use keep_graph=TRUE or retain_graph=TRUE depending on torch version
+       output[i, j]$backward(retain_graph=TRUE)
+       
+       ode_dtheta[i, , j] = as_array(theta$grad)
+       ode_dx[i, , j] = as_array(x$grad[i,])
+     }
+   }
+   
+   list(ode_dtheta = ode_dtheta, ode_dx = ode_dx)
+ }
\end{Sinput}

The correctness of the derivative calculations can be confirmed by comparing the output of \code{hes1logmodelDtheta(theta, x, tvec)} or \code{hes1logmodelDx(theta, x, tvec)} with the output of 
\code{ode_autograd(hes1logmodelODE_torch, theta, x, tvec)} for any given \code{theta}, \code{x}, and \code{tvec}. It is important to note, however, that the computation speed of the autograd version \code{ode_autograd(hes1logmodelODE_torch, theta, x, tvec)} is significantly slower than that of the hand-coded derivatives \code{hes1logmodelDtheta(theta, x, tvec)} or \code{hes1logmodelDx(theta, x, tvec)}.

To utilize \pkg{MAGI} with autograd, we can now proceed to define the \code{odeModel} list containing the three ODE model functions and the parameter bounds:

\begin{Sinput}
R> hes1logmodel <- list(
+   fOde = hes1logmodelODE,
+   fOdeDx = function(theta, x, tvec) 
+     ode_autograd(hes1logmodelODE_torch, theta, x, tvec)$ode_dx,
+   fOdeDtheta = function(theta, x, tvec) 
+     ode_autograd(hes1logmodelODE_torch, theta, x, tvec)$ode_dtheta,
+   thetaLowerBound = rep(0, 7),
+   thetaUpperBound = rep(Inf, 7)
+ )
\end{Sinput}

Note that the original \proglang{R} array implementation \code{hes1logmodelODE} must still be passed to the \code{MagiSolver} function, as \pkg{MAGI} does not currently support direct use of torch tensors. A complete \proglang{R} script that demonstrates this approach of using \pkg{MAGI} with autograd is provided in the replication package.

Although autograd offers a convenient method for calculating derivative information, its computational speed is slower than hand-coded analytical gradients. For optimal performance, we recommend using hand-coded analytical gradients, as discussed in the main text of this paper.

\section[Other covariance functions available in MAGI]{Other covariance functions available in \pkg{MAGI}} \label{app:covariance}

As discussed in Section \ref{subsec:phi}, the default and recommended GP covariance function for use in \pkg{MAGI} is the Matern (Equation \ref{eq:matern}) with $\nu = 2.01$. Several other covariance kernels are also available in the package, which include some of the common choices discussed in Chapter 4 of \cite{williams2006gaussian}. Their specification and features are presented below. In each case, $r$ is the absolute difference between two time points and $\bm{\phi}$ are the hyper-parameters for the kernel; larger values of $\phi_1$ favor curves with higher variance, and larger values of $\phi_2$ favor curves with more time-dependence between nearby time points. They may be selected for use in \code{gpsmoothing} and \code{MagiSolver} by specifying their corresponding \code{kerneltype} string.

\begin{itemize}
	\item Radial basis function or squared exponential (\code{kerneltype = "rbf"}):
\begin{equation*}
\mathcal{K}(r) = \phi_1 \exp \left(-\frac{r^2}{2 \phi_2^2} \right)
\end{equation*}
This is an infinitely differentiable kernel, and hence is associated with very smooth GPs. It may be too smooth to adequately model many physical processes \citep[p.~83,][]{williams2006gaussian}.
	\item Matern with $\nu = 5/2$ (\code{kerneltype = "matern"}):
\begin{equation}\label{eq:matern52}
\mathcal{K}(r) = \phi_1 \left( 1+\frac{\sqrt{5}r}{\phi_2}+\frac{5r^2}{3\phi_2^2} \right) \exp \left(-\frac{\sqrt{5}r}{\phi_2}\right)
\end{equation}	
Equation~\ref{eq:matern52} is a simplification of Equation~\ref{eq:matern} in the special case $\nu = 5/2$. It is faster to compute than $\nu = 2.01$ but has a stronger smoothness assumption, which limits its applicability to systems that are known to have smooth curves.
	\item Compact kernel (\code{kerneltype = "compact1"}): 
\begin{equation*}
\mathcal{K}(r) = \phi_1 \left[\max \left(1-\frac{r}{\phi_2}, 0 \right)\right]^4 \left(\frac{4r}{\phi_2} + 1\right)
\end{equation*}	
This is a kernel with compact support, i.e., the covariance decays to zero for $r \ge \phi_2$, so that points more than $\phi_2$ apart are \textit{a priori} independent. Its polynomial construction also tends to favor smooth curves.

	\item Periodic Matern (\code{kerneltype = "periodicMatern"}):
	
	This follows a ``time warping'' idea to create a non-stationary kernel that could be appropriate for systems that are known to be exactly periodic. 	
	Define the ``time warping'' transformation $r' = |2 \sin(r \pi / \phi_3) |$, where $\phi_3$ is the periodicity parameter. Then the covariance is given by $\mathcal{K}(r')$ using Equation~\ref{eq:matern52}.
		
\end{itemize}

\end{appendix}


\end{document}